\documentclass[letterpaper]{article} 
\usepackage{aaai2026} 
\usepackage{times} 
\usepackage{helvet} 
\usepackage{courier} 
\usepackage[hyphens]{url} 
\usepackage{graphicx} 
\usepackage{natbib} 
\usepackage{caption} 
\usepackage[noend,ruled,linesnumbered,vlined]{algorithm2e}
\usepackage{tikz}
\usetikzlibrary{positioning}
\usetikzlibrary{arrows.meta}

\usepackage{pgfplots}
\pgfplotsset{compat=1.18}

\SetKwInput{Input}{Input} 
\SetKwInput{Output}{Output} 
\SetKwInput{Parameter}{Parameter} 

\usepackage{amsmath}
\usepackage{amsfonts}
\usepackage{appendix}
\usepackage{cleveref}
\newcommand{\RR}{{\bf R}}

\newcommand{\ZZ}{{\mathbb Z}}

\newcommand{\FF}{{\mathbb F}}

\newcommand{\cP}{{\cal P}}
\newcommand{\cW}{{\cal W}}

\newcommand{\cA}{{\cal A}}

\newcommand{\ignore}[1]{}
\newcommand{\JOURNAL}[1]{}

\newtheorem{theorem}{Theorem}[section]

\newtheorem{lemma}[theorem]{Lemma}

\newtheorem{definition}[theorem]{Definition}

\newcommand{\be}{\begin{equation}}
\newcommand{\ee}{\end{equation}}

\usepackage{newfloat}
\usepackage{listings}
\DeclareCaptionStyle{ruled}{labelfont=normalfont,labelsep=colon,strut=off} 
\title{Truth, Justice, and Secrecy: Cake Cutting Under Privacy Constraints}

\author{
Yaron Salman\textsuperscript{\rm 1}\thanks{The first two authors contributed equally to this work.}\!,
Tamir Tassa\textsuperscript{\rm 1}\footnotemark[1],
Omer Lev\textsuperscript{\rm 2},
Roie Zivan\textsuperscript{\rm 2}
}

\affiliations{
\textsuperscript{\rm 1}The Open University of Israel\\
\textsuperscript{\rm 2}Ben-Gurion University of the Negev\\
sayaron8@post.openu.ac.il, tamirta@openu.ac.il, omerlev@bgu.ac.il, zivanr@bgu.ac.il
}

\begin{document}

\maketitle

\begin{abstract}
Cake-cutting algorithms, which aim to fairly allocate a continuous resource based on individual agent preferences, have seen significant progress over the past two decades. Much of the research has concentrated on fairness, with comparatively less attention given to other important aspects.
Chen et al. (2010) introduced an algorithm that, in addition to ensuring fairness, was strategyproof---meaning agents had no incentive to misreport their valuations.
However, even in the absence of strategic incentives to misreport, agents may still hesitate to reveal their true preferences due to privacy concerns (e.g., when allocating advertising time between firms, revealing preferences could inadvertently expose planned marketing strategies or product launch timelines).
In this work, we extend the strategyproof algorithm of Chen et al. by introducing a privacy-preserving dimension. To the best of our knowledge, we present the first private cake-cutting protocol, and, in addition, this protocol is also envy-free and strategyproof. Our approach replaces the algorithm’s centralized computation with a novel adaptation of cryptographic techniques, enabling privacy without compromising fairness or strategyproofness. Thus, our protocol encourages agents to report their true preferences not only because they are not incentivized to lie, but also because they are protected from having their preferences exposed.
\end{abstract}

\medskip
\noindent\textbf{Note.} This is the full version of our paper published in the Proceedings of AAAI 2026.


\section{Introduction}

For millennia, people have grappled with the challenge of dividing resources---whether land, water, loot, or other valuable commodities---in a way that is seen as fair. The modern formalism of this problem---\emph{cake-cutting}---has existed for only a little over 80 years, yet it has found wide application across domains where multiple agents (human or otherwise) assign different values to different parts of a divisible resource. From scheduling shared computer time to allocating energy from power sources, such problems are all instances of cake-cutting, where the goal is to divide the resource fairly among the agents involved.

Naturally, the issue of what is ``fair'' has been widely discussed, with the most basic concepts being \emph{proportionality}, in which each agent gets at least $\frac{1}{n}$ of the value of the whole (assuming $n$ agents overall). A stronger concept is an \emph{envy-free} allocation, in which each agent is most satisfied with its allocation and
has no reason to switch and get a different agent's allocation. There have been many further fairness concepts, and various algorithms have been suggested, aiming for strong guarantees for as wide as possible family of valuation functions, with as low as possible computational complexity \cite{BramsT96,RobertsonW98,BCELP16}.

Although a major open problem in the field---envy-free cake-cutting for $n>3$ agents---was finally solved in the last decade \cite{AM16}, there are many other dimensions to the problem. As with voting, an algorithm guaranteeing many properties may be worthless if agents do not report to the algorithm their true valuations. Envy-free allocations of misreported valuations might not be, in practice, envy-free for the genuine, true valuations. Hence, strategyproof algorithms, in which agents are never hurt by being truthful, are highly desirable.

Work on strategyproof algorithms began with \citet{ChenLPP10,ChenLPP13}, showing a deterministic algorithm for piecewise uniform valuations (e.g., each agent has times when it can use the computer and times when it cannot), which we denote in this paper as \textsc{CC\_puv}; and a randomized algorithm for piecewise linear valuations. Since that work, there has been additional work on strategyproof algorithms \cite{MT10,MN12,ML17,BCHTW17,BHS20}. 
Some work explicitly focused on piecewise uniform, piecewise constant and piecewise linear valuations, as it became quite clear that a deterministic strategyproof algorithm may be limited to a subset of these, with a few additional constraints \cite{AY14,BST23}.

However, while strategyproofness is a desired property, it does not, in fact, completely solve the issue of convincing agents to be truthful and thus obtain the algorithm's guarantees. In many cases, agents do not wish to share with others their valuations for reasons that can be regulatory
(e.g., data protection laws in healthcare may restrict revealing valuations tied to patient information), personal
(e.g., when dividing time to use a game console, one may not wish to reveal to their employer that they are playing during work hours) or due to external effects (e.g., telecom companies bidding on spectrum fear competitors will glean from their bids their future plans and directions). In such cases, agents may withhold their preferences even from a strategyproof algorithm if its operation exposes those preferences to other agents, despite this reducing the algorithm’s effectiveness (e.g., invalidating envy-freeness guarantees).  

In this paper, we propose what we believe to be the \textbf{first private cake-cutting protocol}, named \textsc{PP\_CC\_puv}. This protocol is a privacy-preserving variant of \citet{ChenLPP10,ChenLPP13}'s algorithm and retains its properties of envy-freeness, Pareto-optimality, and strategyproofness (for piecewise uniform valuations). In fact, this is the first protocol that genuinely allows agents to reveal their true valuations without hesitation. It accomplishes this by using secret sharing and secure multiparty computation (MPC) techniques. Secret sharing is used to hide information such as the preferences of agents, the interim results of the computation performed by the algorithm, and the final allocations, while MPC is used to securely perform the computation of the algorithm without exposing the agents' private information.

Achieving this private variant is not simply a matter of inserting standard cryptographic replacements into the original algorithm; it requires novel adaptations of cryptographic techniques. In addition, the algorithm itself had to be restructured. For example, the algorithm of \citet{ChenLPP10,ChenLPP13} constructs a different graph in each iteration, computes a maximum flow in it, and then derives allocations from it. However, the structure of those graphs in each of the iterations might leak sensitive valuation information. To address this, our protocol replaces the dynamic graph with a fixed graph used across all iterations, with edge weights computed cryptographically. Despite this additional complexity, the private variant incurs only $\mathcal{O}(1)$ computational overhead and at most $\mathcal{O}(n^{2})$ additional communication cost.

Beyond privacy, our protocol also allows for the removal of a central coordinating agent to run the algorithm as the participating agents can run it for themselves. This flexibility, requiring no algorithm infrastructure besides the agents, allows cake-cutting algorithms to run on much wider and ad-hoc scenarios, from allocating work shifts between employees (who do not trust the boss to be fair) to computer resource allocation without a pre-existing time-allocation daemon running in the background.

While this is the first private cake-cutting algorithm, we believe that the techniques and ideas it introduces can be extended to “privatize” a broader range of cake-cutting algorithms. Privacy is important in many settings, and we believe that incorporating privacy into the suite of desiderata for resource allocation algorithms is increasingly vital. This is true even when strategyproofness is unnecessary (e.g., for non-strategic agents), as agents may still be reluctant to disclose their valuations to others. 
A natural direction for future work is to extend these ideas to general-purpose cake-cutting algorithms, which are applicable to large classes of valuation functions and stand to benefit significantly from privacy-preserving adaptations that could expand their practical usability.

\section{Cryptographic Background}
Here we briefly review the two fundamental cryptographic techniques underpinning our protocol---secret sharing
(Section~\ref{sss}) 
and multiparty computation (MPC) over secret shares (Section~\ref{MPC})---and describe the security model and privacy guarantees (Section~\ref{model}).

\subsection{Threshold Secret Sharing}\label{sss}
Secret sharing schemes \cite{Shamir79} are protocols that enable distributing a secret scalar $s$ among a set of parties, $P_1,\ldots,P_n$.
Each party, $P_i$, $ i \in [n]$\footnote{For any integer $n$ we let $[n]$ denote the set $\{1,\ldots,n\}$.}, gets a random value $[[s]]_i$, called {\em a share}, 
so that some subsets of those shares enable the reconstruction of $s$, while each of the other subsets of shares reveals no information on $s$. 
In its most basic form, called {\em Threshold Secret Sharing}, there is a threshold value $t \leq n$, and then a subset of shares enables the reconstruction of $s$ iff its size is at least $t$.

Shamir's $t$-out-of-$n$ threshold secret sharing scheme \cite{Shamir79} operates over a finite field $\FF_p$, where $p>n$ is a prime sufficiently large so that all possible secrets may be represented in $\FF_p$.
It has two procedures---Share and Reconstruct:

$\bullet$ $\text{Share}_{t,n}(s)$.
The procedure samples a uniformly random polynomial $f(\cdot)$ over $\FF_p$, of degree at most $t-1$, where the constant term is the secret $s$. The procedure outputs $n$ values, $[[s]]_i=f(i)$, $i \in [n]$, where $[[s]]_i$ is the share given to $P_i$. The entire set of shares, denoted $[[s]]:=\{ [[s]]_i : i \in [n] \}$, is called a $(t,n)$-sharing of $s$.

$\bullet$ $\text{Reconstruct}_{t}([[s]])$. The procedure is given any selection of $t$ shares out of $[[s]]$---the $(t,n)$-sharing of $s$.
It then interpolates a polynomial $f(\cdot)$ of degree at most $t-1$ using the given points and outputs $s=f(0)$.
Any selection of $t$ shares
will yield the secret $s$, as $t$ points determine a unique polynomial of degree at most $t-1$.
On the other hand, any selection of $t-1$ shares or less reveals nothing about the secret $s$.

Hereinafter, we set the threshold to be 
\be t := \lfloor (n+1)/2 \rfloor \,.\label{tdef}\ee
Namely, to reconstruct the secret, at least half of the parties must collaborate. Hence, if the set of $n$ parties has an honest majority,
in the sense that more than half of them act honestly, without trying to collude to reconstruct the secret illicitly,
the shared secret will remain fully protected. 

\JOURNAL{
In what follows, we shall use the following terminology and notations.
Let $s$ be a secret known to some party $P_i$, $i \in [n]$. Then if $P_i$ performs the procedure $\text{Share}_{t,n}(s)$, we will simply say that $P_i$ distributes a $(t,n)$-sharing of $s$.

If the parties have a $(t,n)$-sharing $[[s]]$ of some secret $s$ and they wish to let one of them, say $P_i$, reconstruct the secret $s$, then at least $t-1$ parties would send their shares to $P_i$ who will proceed to apply the Reconstruct procedure on the $t$ shares it has.
}

\subsection{Multiparty Computation Over Secret Shares}\label{MPC}
Let $u$ and $v$ be two secret values in the field $\FF_p$, and assume that $P_1,\ldots,P_n$ hold $(t,n)$-sharings of them, denoted
$[[u]]=\{[[u]]_i: i \in [n]\}$ and
$[[v]]=\{[[v]]_i: i \in [n]\}$.

We are concerned here with performing secure computations over those shared values in the following sense: if $G(\cdot,\cdot)$ is a publicly known function, the goal is to use given $(t,n)$-sharings $[[u]]$ and $[[v]]$ in order to compute a $(t,n)$-sharing of $G(u,v)$, without learning any information on $u$, $v$, or $G(u,v)$. We consider here basic functions that enable performing a wide range of more elaborate computations.

{\bf Affine combinations.} 
Let $\alpha,\beta,\gamma\in \FF_p$ be three values publicly known to all. Then 
\[ [[\alpha]]+\beta[[u]]+\gamma[[v]] := \{ \alpha+\beta[[u]]_i + \gamma[[v]]_i : i \in [n]\} \]
is a proper $(t,n)$-sharing of $w:=\alpha+\beta u + \gamma v$, thanks to the affinity of secret sharing.
By writing
$ [[w]] \gets [[\alpha]]+\beta[[u]]+\gamma[[v]] $
we mean that each party $P_i$, $i \in [n]$, sets
$[[w]]_i \gets \alpha+\beta[[u]]_i + \gamma[[v]]_i $, so that now the parties hold a $(t,n)$-sharing of
$w=\alpha+\beta u + \gamma v$, without needing to interact or to perform any further polynomial computations. 

By choosing $\beta = \gamma = 0$, it follows that for any public value $\alpha$ in the field, the set $\{ [[\alpha]]_i = \alpha : i \in [n] \}$ constitutes a valid $(t,n)$-secret sharing of $\alpha$.

{\bf Multiplications.}
A secure multiplication protocol,
$ [[w]] \gets 
[[ u]] \cdot[[ v]] $,
takes $(t,n)$-sharings of $u$ and $v$ and computes a $(t,n)$-sharing of $w=u\cdot v$, without revealing to the parties any information on $u$, $v$, or $w=uv$.
\citet{DN07} designed such a secure multiplication protocol, which was later improved by \citet{ChidaGHIKLN18}.

\JOURNAL{
In what follows we shall use the shortened notation $[[w]] \gets [[ u]] \cdot[[ v]]$ instead of the notation in Eq. (\ref{sm}).}

{\bf Comparisons.}
Hereinafter, if $\cP$ is a predicate then $1_{\cP}$ is a bit that equals 1 if the predicate $\cP$ holds and equals 0 otherwise. Then a secure comparison protocol, $ [[w]] \gets [[ 1_{u<v} ]] $,
takes $(t,n)$-sharings of $u$ and $v$ and securely computes a $(t,n)$-sharing of $w=1_{u<v}$. \citet{NO07} proposed such a secure comparison protocol.
\JOURNAL{
(In some cases we shall invoke a protocol of the form $[[w]] \gets [[ 1_{u\leq v} ]]$, which is equivalent to $[[w]] \gets [[ 1_{u<v+1} ]]$ (as $u$ and $v$ are integers).
}

{\bf Equality.}
A secure equality testing protocol,
$[[w]] \gets 
[[ 1_{u=0} ]] $,
takes a $(t,n)$-sharing of $u$ and securely computes a $(t,n)$-sharing of $w=1_{u=0}$, see \cite{KoganTG23}. 

{\bf Minimum.}
A secure minimum protocol,
$ [[w]] \gets \min([[u]], [[v]])$,
takes $(t,n)$-sharings of $u$ and $v$, and securely computes a $(t,n)$-sharing of their minimum, $w=\min(u,v) = v + 1_{u<v} \cdot (u-v)$. 
Maxima can be computed similarly by the identity $\max(u,v)=u + 1_{u<v} \cdot (v-u)$.

{\bf OR.}
A secure OR protocol,
$ [[w]] \gets [[ u]] \vee [[ v ]]$,
takes $(t,n)$-sharings of two secret bits, $u$ and $v$, and securely computes a $(t,n)$-sharing of the OR between them, $w=u \vee v = u+v - u \cdot v$. 

{\bf Divisions.} A secure division protocol,
$ \langle [[q]],[[r]] \rangle \gets [[u]]/[[v]]$, 
takes $(t,n)$-sharings of two secret integers, $u$ and $v$, and securely computes $(t,n)$-sharings of the corresponding quotient $q = \lfloor \frac{u}{v} \rfloor $ and remainder $r = u ~\mbox{mod} ~v$. In Appendix A.1 we present such a protocol.

\subsection{Security Model and Privacy Guarantees}\label{model}
Our protocol operates in a standard MPC setting in which the computation is carried out by several semi-honest servers, with an honest majority ensuring that any colluding subset is smaller than half.
Agents may attempt to behave dishonestly, but the protocol incorporates integrity checks that prevent them from submitting valuation data that deviates from the required piecewise-uniform structure.
Privacy is guaranteed in an information-theoretic sense: no party---neither agents nor servers---learns anything beyond the outputs explicitly revealed by the protocol.
In particular, the agents' valuation functions, all intermediate values (including demands, interval selections, and flow computations), and the structure of the underlying decision processes remain completely hidden.
The only information that may be disclosed is the final allocation, which can be revealed either in full or only to the corresponding agents, depending on the chosen visibility mode.

\section{Strategyproof Cake Cutting Algorithm}\label{Sec:prelim}
We begin by presenting \citet{ChenLPP10,ChenLPP13}'s cake-cutting algorithm, which we refer to as \textsc{CC\_puv}—the \underline{C}ake \underline{C}utting algorithm for \underline{p}iecewise \underline{u}niform \underline{v}aluations.

The cake to be divided is modeled as the interval $C=[0,1] \subset \RR$.
The set of agents that seek to divide the cake between them is $\cA:=\{A_i: i \in [n]\}$.
Agent $A_i$ has a valuation function $v_i : C \rightarrow [0,\infty)$ that is a probability density function over $C$; i.e., it is measurable, nonnegative, and $\int_C v_i(x) dx=1$. Given any piece of the cake, $X \subseteq C$, its value to $A_i$ is
$ v_i(X) :=\int_X v_i(x) dx $.

\textsc{CC\_puv} assumes all valuations are piecewise uniform:

\begin{definition}\label{puvdef} A valuation $v: C \rightarrow [0,\infty)$ is called piecewise uniform if there exist $\ell \geq 1$ disjoint intervals, $\{I_j \subseteq C: j \in [ \ell]\}$, such that $v(x)= c$ for all $x \in B:=\bigcup_{j\in [ \ell]} I_j$, for $c = \left( \sum_{j\in [ \ell]} |I_j| \right)^{-1}$, while $v(x)=0$ for all $x \in C \setminus B$.
\end{definition}

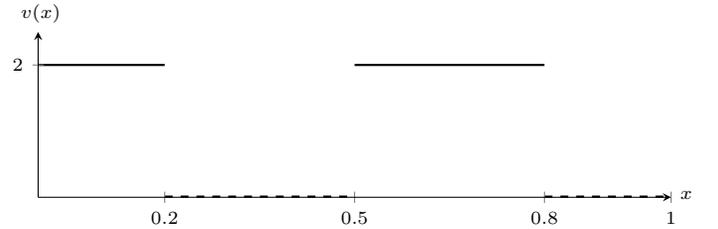
\begin{figure}[h!!!]
\centering
\begin{tikzpicture}
\begin{axis}[
    axis lines=middle,
    xmin=0, xmax=1,
    ymin=0, ymax=2.5,
    xtick={0,0.2,0.5,0.8,1},
    ytick={0,2},
    xlabel={$x$},
    ylabel={$v(x)$},
    xlabel style={anchor=west, font=\scriptsize, yshift=1pt},
    ylabel style={anchor=south, font=\scriptsize, xshift=1pt},
    label style={font=\scriptsize},
    tick label style={font=\scriptsize},
    width=\columnwidth,
    height=2.2cm,
    scale only axis,
    clip=false,
    axis on top,
    tick style={thin},
    every axis plot/.append style={thick},
    enlargelimits=false,
    clip mode=individual
]

\addplot[domain=0:0.2, samples=2] {2};
\addplot[domain=0.5:0.8, samples=2] {2};
\addplot[domain=0.2:0.5, samples=2, dashed] {0.01};
\addplot[domain=0.8:1, samples=2, dashed] {0.01};

\end{axis}
\end{tikzpicture}
\caption{\small A piecewise uniform valuation with $\ell=2$ intervals.}
\label{puv}
\end{figure}

Figure \ref{puv} illustrates such a function with a support consisting of $\ell=2$ intervals.

The end points of the support intervals in $v_i(\cdot)$, over all $i \in [n]$, induce a partition of $C=[0,1]$ into a set of intervals, which we denote by $\cW$.
Given an interval $I \in \cW$ and a subset $S \subseteq \cA$ of agents, $S$ desires $I$ if $I$ is included in the support of the valuation of at least one agent in $S$.
If $X \subseteq \cW$ then $D(S,X)$ is the subset of all intervals in $X$ that are desired by $S$, and
$\text{avg}(S, X) := \text{Len}(D(S, X)) / |S|$ is the average demand of $S$ on $X$. (If $Y$ is a set of disjoint intervals, then $\text{Len}(Y)$ is the sum of lengths of those intervals.) 

The deterministic algorithm of \citet{ChenLPP10}, \textsc{CC\_puv}, operates iteratively. It starts with $X = \cW$ and $S = \cA$.
It then identifies a subset of agents $S' \subseteq S$ (from the current agent set $S$) with the smallest average demand on the current set of intervals $X$. 
The algorithm proceeds to perform an \emph{exact allocation} of $D(S', X)$ among $S'$, giving each agent in $S'$ an equal share of size $\text{avg}(S', X)$ out of those intervals in $X$ that it desires. 
The algorithm computes these allocations by solving a maximum-flow problem on a capacitated directed graph that encodes the desirability relationships between the intervals in $X$ and the agents in $S'$. A detailed explanation of this computation is provided in Section~\ref{Step3}. The procedure then recurses on the remaining agents and intervals, specifically on $S \setminus S'$ and $X \setminus D(S', X)$, until all agents have been served—that is, until $S = \emptyset$.

\section{Privacy-Preserving Cake-Cutting Protocol}
In this section, we present our privacy-preserving cake-cutting protocol, which securely implements the cake-cutting algorithm introduced by \citet{ChenLPP10,ChenLPP13}.
We begin with a high-level overview of our privacy-preserving protocol, \textsc{PP\_CC\_puv}---the \underline{P}rivacy-\underline{P}reserving variant of \textsc{CC\_puv} (Section~\ref{overview}).
The subsequent sections describe the core computational steps of \textsc{PP\_CC\_puv}.

\subsection{Overview of \textsc{PP\_CC\_puv}}\label{overview}
The main challenge in implementing \textsc{CC\_puv} securely is preserving the privacy of the agents' valuations.
Our privacy-preserving protocol, Protocol~\ref{ppcc} (\textsc{PP\_CC\_puv}), addresses this challenge by having each agent distribute secret shares of their piecewise-uniform valuations.
It then faithfully emulates \textsc{CC\_puv} by executing carefully tailored MPC subprotocols on these secret shares.

\textsc{PP\_CC\_puv} begins with several preparatory steps.
First (line~1), the agents execute a subprotocol called \textsc{SharingPrivateValuations}, which (a) discretizes the private valuations in a lossless manner, (b) enforces integrity checks to prevent dishonest agents from distributing shares that do not correspond to valid valuations, and (c) distributes secret shares of the private valuations (\Cref{preparing}).

Next (lines~2--3), the secret-shared valuation functions are used to partition the cake \(C=[0,1]\) into a collection of intervals, denoted \(\cW\), such that each interval lies entirely within the support of \(v_i\) or entirely outside it, for every \(i\in[n]\).
A subprotocol called \textsc{Intervals} establishes secret shares of various auxiliary data structures, such as indicator bits specifying, for each agent and interval, whether the agent desires that interval (\Cref{sec_split,intsec}).

At this point, \textsc{PP\_CC\_puv} enters its core iterative phase (lines~5--8).
In each iteration, it executes a subprotocol called \textsc{IterativeAllocation} that selects a subset of yet-unserved agents, according to the same selection criterion used by \textsc{CC\_puv}.
The key challenge here is to keep the protocol oblivious to which agents have already been served and which intervals remain available (\Cref{basic_alg}).
Once the appropriate subset is (obliviously) selected, the allocation of intervals (or partial intervals) to agents is performed by solving a maximum-flow problem on a graph that encodes the relationships between the relevant agents and intervals.
Crucially, the subprotocol \textsc{AssignCakeToSelectedAgents} does this while remaining oblivious to the structure of the underlying graph (\Cref{Step3}).

Once the iterative process is complete, the subprotocol \textsc{FinalServing} scans all intervals and assigns them to the appropriate agents, based on the computation that took place during the iterative phase (line~9).
Depending on the application scenario, the protocol can operate under either restricted visibility, where each agent learns only their own allocated piece, or full visibility, where the complete allocation is disclosed to all agents (\Cref{finalcut}).

\SetAlgoNlRelativeSize{-1}
\SetAlgoNlRelativeSize{0}
\SetAlgoNlRelativeSize{1}
\SetAlgorithmName{Protocol}{}{} 

\SetNlSty{textnormal}{\scriptsize}{} 

\begin{algorithm}[htb!]
\caption{\textsc{PP\_CC\_puv}\label{ppcc}}

\SetAlgoLined
\DontPrintSemicolon

Distribute secret shares of private valuations\;
Split the cake into intervals $\cW$\;
Compute a $\cW$-based representation of the valuations\;
$S \gets [n],~X \gets \cW$\;

\While{$S \neq \emptyset$}{\label{loopStarts}
 Find $S' \subseteq S$ minimizing $\text{avg}(S', X)$ (the average demand of agents in $S'$ over intervals in $X$)\;
 Compute a fair allocation of intervals in $D(S', X)$ (the subset of intervals in $X$ desired by $S'$) to agents in $S'$\;
 $S \gets S \setminus S',~X \gets X \setminus D(S', X)$\;
}

Distribute the intervals in $\cW$ among the agents\;

\end{algorithm}

\subsection{Secret Sharing the Private Valuation Functions}\label{preparing}
Here we describe how the agents secret share their private valuation functions.
Let $\ell_i$ denote the number of intervals in $v_i$'s support, $i \in [n]$ and let $ \ell = \max_{i \in [n]} \ell_i$.
Then the support of $v_i$ can be described by
$$ \mbox{supp}(v_i) = \bigcup_{j=1}^\ell I_{i,j}\,, ~ \mbox{where} ~ I_{i,j}=[a_{i,j},b_{i,j})\,,~ j \in [\ell]\,, ~~~\mbox{and}$$
\be 0 \leq \hskip -0.05cm a_{i,1} \leq b_{i,1} \leq a_{i,2} \leq b_{i,2} \leq \cdots \leq a_{i,\ell} \leq b_{i,\ell} \leq 1\,. \label{boundaries}\ee
By invoking a secure maximum computation (see Section~\ref{MPC}), the agents can compute $\ell$ without disclosing any additional information about $\ell_i$, $i \in [n]$. Then, if $\ell_i<\ell$, agent $A_i$ sets the last $\ell-\ell_i$ intervals in $v_i$'s support to the empty interval $[1,1)$.
Such an inflated description of the valuations guarantees that no information is leaked about the number of intervals in the valuations' supports.

The end points of the intervals in those supports, Eq. (\ref{boundaries}), will be protected by secret sharing. As secret sharing is applied in some finite field, it is essential to convert those real values into integers. To do so, the agents jointly agree upfront on some sufficiently large integer,
$Q$, and then each agent $A_i$, $i \in [n]$, redefines its end points as follows,
$a_{i,j}:= \lfloor a_{i,j} \cdot Q \rceil$ and
$b_{i,j} := \lfloor b_{i,j} \cdot Q \rceil$, where $\lfloor \cdot \rceil$ denotes the closest integer.
By taking $Q=10^d$, where $d$ is the maximum number of nonzero decimal digits after the decimal point among the numbers $\{a_{i,j},b_{i,j}\}_{i \in [n],j\in [\ell]}$, such a discretization is lossless. 
Therefore, henceforth the support of $v_i$ will be described by $2\ell$ integers 
\be 0 \leq \hskip -0.05cm a_{i,1} \hskip -0.05cm \leq
\hskip -0.05cm
b_{i,1} \hskip -0.05cm \leq
\hskip -0.05cm a_{i,2} \hskip -0.05cm \leq
\hskip -0.05cm
b_{i,2} \hskip -0.05cm \leq \cdots \leq a_{i,\ell} \leq b_{i,\ell} \leq Q\,. \label{boundaries_int}\ee
Now, the agents proceed to distribute shares of the $2\ell $ integers in Eq. (\ref{boundaries_int}), for all $i \in [n]$.
Such a procedure must incorporate a computation of $\ell = \max_{i \in [n]} \ell_i$ and of $d$ (as described above), as well as mechanisms to detect possible cheats of dishonest agents.

This is carried out by Subprotocol 4,
termed \textsc{SharingPrivateValuations} (see Appendix A.2).
After its completion, the agents hold $(t,n)$-sharings in $a_{i,j}$ and $b_{i,j}$ for all $i \in [n]$ and $j \in [\ell]$. 

\subsection{Splitting the Cake to Intervals}\label{sec_split}
Consider the multiset $ V := \{a_{i,j},b_{i,j} : i \in [n], j \in [\ell]\} \cup \allowbreak \{0,Q\} $, of size $m = 2 \ell n + 2$,
consisting of all support boundaries of all valuations, as well as the cake's endpoints.
Let $W(1:m)$ be an array that holds the values in $V$ sorted, namely, $W(1) \leq \cdots \leq W(m)$. Then $W$ defines a collection of
$m-1$ disjoint intervals (some of which may be empty) that cover the entire cake:
\be \cW := \big\{ [W(k),W(k+1)) : k \in [m-1] \big \} \,. \label{intervals} \ee
Note that $[W(k),W(k+1)) \subset [0,Q]$, and that the actual interval it represents on $C=[0,1]$ is $[\frac{W(k)}{Q},\frac{W(k+1)}{Q})$.

After completing Subprotocol \textsc{SharingPrivateValuations}, the agents hold $(t,n)$-sharings of all values in $V$. They can proceed to compute a secret sharing of $W(k)$, for all $k \in [m]$, by applying a suitable secure protocol that computes the $k$-th element (counting from the smallest) in a secret-shared dataset, without getting any wiser on the private valuations. Such a protocol is described in \citet{MPCoverSS}.

Let $\mbox{IntervalLen}(k)= W(k+1)-W(k)$ denote the length of the $k$-th interval, $k \in [m-1]$.
Secret shares in those intervals' lengths can be computed by $[[ \mbox{IntervalLen}(k) ]] \gets
[[ W(k+1) ]] - [[ W(k) ]] $.
In addition, the protocol maintains secret sharings of the bits $\mbox{IntervalAvailable}(k)$, $k \in [m-1]$, that indicate whether the $k$-th interval has not yet been served to any agent.
Initially, $ [[ \mbox{IntervalAvailable}(k) ]] \gets
[[1]]$. Once the $k$-th interval is served, the protocol updates the shares of $\mbox{IntervalAvailable}(k)$ to reflect the bit update from 1 to 0, in an {\em oblivious manner}, namely, without knowing that such an update actually took place and to whom that interval was served. 

\subsection{Encoding the Valuations by Binary Vectors}\label{intsec}
The intervals in $\cW$, Eq. (\ref{intervals}), can be used to encode the private valuations by a bit array, $\mbox{IntervalDesired}(1:n,1:m-1)$, as follows: $\mbox{IntervalDesired}(i,k)=1$ iff the $k$-th interval in $\cW$ is desired by $A_i$. 
Such desirability holds iff that interval is of positive length and it is contained in $v_i$'s support. Hence,
\be
\begin{aligned}
\mbox{IntervalDesired}(i,k) 
= \left(1 - 1_{\text{IntervalLen}(k)=0} \right) \hskip 1cm \\
\hskip .2cm \cdot \bigvee_{j \in [\ell]} \left( 1_{a_{i,j} \leq W(k)} \cdot 1_{W(k+1) \leq b_{i,j}} \right) \,,
\end{aligned}
\label{postp}
\ee
because $[W(k), W(k+1)) \subseteq \mbox{supp}(v_i)$ iff $[W(k), W(k+1)) \subseteq [a_{i,j},b_{i,j})$ for some $j \in [\ell]$.

The agents may compute secret sharings of $\mbox{IntervalDesired}(i,k)$ for all $i \in [n]$ and $k \in [m-1]$, from the secret sharings that they already hold in $a_{i,j},b_{i,j}$, $W(k)$, and $\text{IntervalLen}(k)$, using the secure MPC protocols for comparison, multiplication, OR, and equality,
as discussed in Section \ref{MPC}.

\JOURNAL{I deleted here a lot of text that appears in the snapshot before AAAI}

Subprotocol 6
(\textsc{Intervals}) in Appendix A.4 
summarizes the interval computations as discussed in Sections \ref{sec_split} and \ref{intsec}. It computes the secret sharings of the arrays $W$, IntervalLen, and IntervalDesired, and initializes the secret sharing of the array IntervalAvailable. 
In addition, it initializes the secret shares of two additional secret arrays that will be used later on: AllocationDenominator$(1:n)$, and
$ \mbox{IntervalAllocation}(1:n,1:m-1)$. 
Subprotocol \textsc{Intervals} initializes their secret shares to zero. They will be kept secret-shared, and only after all agents have been served they will be used to infer the portion of each of the intervals that would go to any specific agent. Specifically, the fraction
\be \mbox{Portion}(i,k):= \frac{\textnormal{IntervalAllocation}(i,k)}{\mbox{AllocationDenominator}(i)} \label{fraction}\ee
will equal the portion of the $k$-th interval allocated to agent $A_i$, $i \in [n]$, $k \in [m-1]$.
The update of those arrays will be explained in due time (Section \ref{Step3}).

\subsection{Selecting a Subset of Agents to Serve}\label{basic_alg}
As explained in Section \ref{Sec:prelim}, the algorithm \textsc{CC\_puv} functions through an iterative process. It maintains two variables:
(1) a subset $S \subseteq [n]$ that represents the agents that have yet to be served a piece of cake; and (b) a subset $X \subseteq \cW $ of cake intervals that still have not been served to anyone.

Initially, $S = [n]$ and
$X = \cW$. 
In each iteration, the algorithm selects $S' \subseteq S$ that minimizes
\be \mbox{avg}(S',X):= \mbox{Len}(S',X)/|S'| \,, \label{avgdef}\ee
where 
$\mbox{Len}(S',X)$ is the sum of lengths of all intervals in $X$ that are desired by at least one agent in $S'$. As can be readily verified, it is given by
\be
\begin{split}
\mbox{Len}(S',X) 
= \sum_{k \in [m-1]} \mbox{IntervalAvailable}(k) \hskip 1.3cm \\
\quad \quad \cdot \left( \bigvee_{i \in S'} \mbox{IntervalDesired}(i,k) \right) \cdot \mbox{IntervalLen}(k)
\end{split}
\label{lendef}
\ee
\JOURNAL{
Indeed, the sum on the right hand side of Eq. (\ref{lendef}) goes over all intervals in $\cW$; the first multiplicand in each addend makes sure that we consider only intervals in $X$; the second multiplicand ascertains that we consider only intervals that are desired by at least one agent in $S'$; and the third multiplicand is the length of the corresponding interval.
}

For privacy considerations, it is necessary to protect the identity of agents who are being served in each iteration. However, at the same time, it is necessary to consider in each iteration only subsets $S'$ that consist entirely of agents who have not yet been served. To achieve this while keeping the served agents’ identities hidden, we introduce the bit array
$\mbox{AgentsServed}(1:n)$,
in which the $i$-th bit is initially set to 0 and is updated to 1 once $A_i$ has been served.
By keeping that array $(t,n)$-secret-shared,
the identities of the agents being served in each iteration remains secret.

Subprotocol 7,
termed \textsc{IterativeAllocation} (see Appendix A.5), carries out the iterative allocation procedure. In each iteration, it searches for a subset \( S' \) that minimizes \( \mbox{avg}(S', X) \), and then allocates to each agent in \( S' \) its portion of the intervals in \( X \) that it desires.
We proceed to survey its main features.

In each of its iterations, subprotocol \textsc{IterativeAllocation} finds a non-empty subset $S' \subseteq [n]$ that minimizes $\mbox{avg}(S',X)$, Eq. (\ref{avgdef}). In order to force the subprotocol to identify a minimizing subset only from among the legal subsets $S'$---those that consist of agents that have not yet been served---we introduce the marker
\be h(S') := \sum_{i \in S'} \mbox{AgentsServed}(i) \,.
\label{hdef}\ee
A subset $S' \subseteq [n]$ is legal iff $h(S')=0$. Define
\be 
\begin{split}
\mbox{Len}^*(S',X) := \mbox{Len}(S',X) \cdot 1_{h(S')=0} \hskip 2cm \\ +n(Q+1) \cdot (1 - 1_{h(S')=0})\,. 
\end{split}
 \label{LenStar}
\ee
The value $\mbox{Len}^*(S',X)$ equals
$\mbox{Len}(S',X)$ for legal subsets $S'$, while for illegal subsets it equals $n(Q+1)$. Hence,
\be \mbox{avg}^*(S',X):= \mbox{Len}^*(S',X)/|S'| \,, \label{avgstar}\ee
equals
$\mbox{avg}(S',X)$ (Eq. (\ref{avgdef})) for legal subsets $S'$, while it is at least $Q+1$ for illegal subsets. Since for all legal subsets $\mbox{avg}(S',X)$ is at most $Q$, any subset $S'$ that minimizes 
$\mbox{avg}^*(S',X)$ would be legal.
Hence, Subprotocol \textsc{IterativeAllocation}
finds a non-empty subset $S' \subseteq [n]$ that minimizes 
$\mbox{avg}^*(S',X)$, using Eqs. (\ref{lendef})--(\ref{avgstar}) and the basic secure MPC protocols described in Section \ref{MPC}. 
Then, it
calls subprotocol
\textsc{AssignCakeToSelectedAgents} (which we discuss below) that assigns cake pieces to the agents in the selected subset $S'$. Finally, Subprotocol \textsc{IterativeAllocation}
updates the secret shares of AgentsServed, according to the identities of the agents in $S'$ who were just served, as well as the secret shares of IntervalAvailable, according to the intervals that were served in this iteration. Those updates are carried out in an oblivious manner, namely, without revealing the identities of those agents nor the indices of the intervals that were served to them.

\subsection{Allocating Cake Pieces to the Agents in the Selected Subset}\label{Step3}
After identifying a subset of agents $S'$ that minimizes the average demand over 
the available set of intervals 
$X$, algorithm \textsc{CC\_puv} \cite{ChenLPP10} computes a fair allocation of the desired intervals in $X$ among the agents in $S'$. 

\begin{figure}[h!!!]
\centering
\scalebox{0.75}{ 
\begin{tikzpicture}[->,>=stealth,thick,node distance=1.2cm and 1.2cm, every node/.style={scale=0.85}]
 \node (A1) at (0,2.4) {$\mbox{Interval}_1$};
 \node (A2) at (0,1.6) {$\mbox{Interval}_2$};
 \node (A3) at (0,0.8) {$\mbox{Interval}_3$};
 \node (A4) at (0,0)   {$\mbox{Interval}_4$};
 \node (A5) at (0,-0.8) {$\mbox{Interval}_5$};


 \node (P1) [right=2.5cm of A2] {$A_1$};
 \node (P2) [right=2.5cm of A4] {$A_3$};
 \node (P3) [right=2.5cm of A3] {$A_2$};

 \node (s) [left=5.5cm of P2, yshift=0.6cm] {$\mbox{source}$};
 \node (t) [right=2.5cm of P2, yshift=0.6cm] {$\mbox{target}$};

 \foreach \i in {1,...,5} {
   \draw (s) -- (A\i.west);
 }

 \draw (A1.east) -- (P1.west);
 \draw (A2.east) -- (P1.west);
 \draw (A3.east) -- (P1.west);
 \draw (A2.east) -- (P2.west);
 \draw (A3.east) -- (P2.west);
 \draw (A4.east) -- (P2.west);
 \draw (A1.east) -- (P3.west);
 \draw (A4.east) -- (P3.west); 

 \foreach \i in {1,2,3} {
   \draw (P\i) -- (t);
 }
\end{tikzpicture}
}
\caption{An example of a flow graph $G(S',X)$}
\label{fig:graph}
\end{figure}
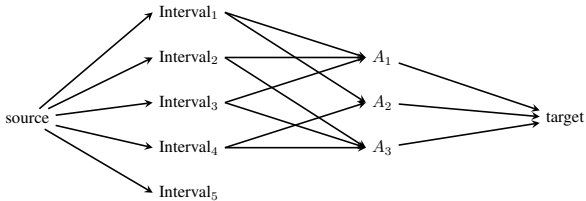

To do so, it constructs a four-layer directed graph $G(S',X)$: layer 1 has a source node; layer 2 has a node for each interval in $X$; layer 3 has a node for each agent in $S'$; and layer 4 has a target node. The graph has an edge from the source node to each of the nodes in layer 2, with a capacity that equals the length of the corresponding interval.
There is an edge from each node in layer 2 to each node in layer 3 that corresponds to an agent that desires that cake interval; the capacity of such an edge is the length of the interval from which it emerges. Finally, it has an edge from each node in layer 3 to the target node with a capacity that equals $\mbox{avg}(S',X)$, Eq. (\ref{avgdef}). 
Figure \ref{fig:graph} illustrates the graph $G(S',X)$ for a set $X$ of 5 intervals and a set $S'$ of 3 agents, who desire 4 out of the 5 available intervals in $X$. 

The graph has a maximum flow of value \(\mbox{Len}(S', X)\). This flow fully saturates the edges between layers 1 and 2 corresponding to intervals in \(X\) desired by \(S'\), while all other edges between these layers carry zero flow. The flow across the edges from layer 2 to layer 3 determines a fair allocation of the \(S'\)-desired intervals in \(X\) among the agents in \(S'\). Finally, all edges between layers 3 and 4 are saturated, ensuring that
each agent in $S'$ gets cake pieces of which the sum of lengths is exactly $\mbox{avg}(S',X)= \mbox{Len}(S',X)/|S'|$.

Subprotocol 9
(\textsc{AssignCakeToSelectedAgents}), 
see Appendix A.6,
emulates the computation described above while addressing a significant algorithmic challenge: how can one compute a maximum flow in a graph that must remain secret? The graph $G(S', X)$ inherently reveals the sets $S'$ and $X$, as well as the desirability relationships between them, and therefore cannot be explicitly constructed. To resolve this challenge, the subprotocol operates on a fixed, publicly known four-layer graph, where layer 2 contains a node for each interval in $\mathcal{W}$ and layer 3 contains a node for each agent in $\mathcal{A}$. Although the structure of this graph is static and public (since $|\mathcal{W}| = m - 1$ and $|\mathcal{A}| = n$), the edge capacities vary across iterations. These capacities are carefully defined to match those in the underlying algorithm \textsc{CC\_puv} for edges between intervals in $X$ and agents in $S'$, while all other capacities are set to zero. Crucially, the capacities are secret-shared to conceal the topology of the actual subgraph, thereby preserving the privacy of $S'$, $X$, and the desirability relationships between agents and intervals.

The maximum flow that Subprotocol \textsc{AssignCakeToSelectedAgents} computes determines the allocation of (a subset of the) intervals to (a subset of the) agents. Those allocations cannot be disclosed at this stage. Instead, they must be registered and disclosed together with all allocations from all iterations only after the iterative allocation process 
terminates. To that end, we use the secret-shared arrays
$ \mbox{IntervalAllocation}(1:n,1:m-1)$ and AllocationDenominator$(1:n)$, that were initialized in Subprotocol \textsc{Intervals} (see Section \ref{intsec}).
AllocationDenominator$(i)$ registers the size of the subset $S'$ to which $A_i$ belonged when it was served, while $ \mbox{IntervalAllocation}(i,k)$ registers the length of the part of the $k$-th interval allocated to $A_i$, multiplied by AllocationDenominator$(i)$ (so that, eventually,
the portion of the $k$-th interval allocated to $A_i$ is the fraction in Eq. (\ref{fraction})). After Subprotocol \textsc{AssignCakeToSelectedAgents} finds a maximum flow, it updates the secret shares of the relevant entries in those two arrays. 
For a comprehensive account, see Appendix A.6.


\JOURNAL{see line 23 in Subprotocol \ref{sffa}.
Since the value $b_{k,i}$ will be nonzero at most in one of the iterations, 
$\mbox{IntervalAllocation}(i,k)$ will equal, eventually, the part of the $k$-th interval that was allocated to $A_i$, multiplied by the size of the subset $S'$ to which $A_i$ belonged (that is $\mbox{SizeBestSubset}$, see Eq. (\ref{abcij})).
In order to get rid of the SizeBestSubset multiplier, we register its value in AllocationDenominator$(i)$, see line 21 in the subprotocol.
Since, for each $i \in [n]$, the bit $\mbox{BestSubset}(i)$ will be 1 only in one of the iterations, and in that iteration SizeBestSubset equals the size of the subset $S'$ in which $A_i$ was a member, the final value of
$\mbox{AllocationDenominator}(i)$ will be the correct denominator by which all
values $\mbox{IntervalAllocation}(i,k)$, $k \in [m-1]$, should be divided.}

\subsection{Cutting the Cake}\label{finalcut}
Once Subprotocol \textsc{IterativeAllocation} terminates, it is necessary to compute the values $\text{Portion}(i,k)$ (Eq.~(\ref{fraction})) that specify the fraction of the $k$-th interval allocated to agent $A_i$, for all $k \in [m-1]$ and $i \in [n]$. Specifically, if $\text{Portion}(i,k) = \alpha$, then $A_i$ receives a subinterval of length $\frac{\alpha}{Q}$ from the interval $[\frac{W(k)}{Q}, \frac{W(k+1)}{Q}) \subset [0,1)$ (see Eq.~(\ref{intervals}) and the accompanying discussion). Subprotocol 11,
referred to as \textsc{FinalServing} (see Appendix A.7), performs this post-processing step by cutting the cake and assigning each agent their corresponding portion. 

Let \(g(k)\) denote the number of agents that were allocated some non-empty part of the $k$-th interval, \(k \in [m-1]\).
Intervals for which \(g(k) = 1\) are considered \emph{exclusive}, meaning they are fully assigned to a single agent. Intervals for which \(g(k) \neq 1\) are \emph{non-exclusive}; they are either split among multiple agents or unassigned altogether. 
The subprotocol records this classification in the secret-shared array, \(\textnormal{ExclusiveInterval}(1:m-1)\), in which the $k$-th entry equals the index of the single agent that was assigned the entire \(k\)-th interval, when \(g(k) = 1\), or 0 otherwise.

After completing the initial loop to determine the type of each interval, the subprotocol proceeds with a second loop over all intervals, $k \in [m-1]$, that assigns each interval to the appropriate agent or agents.
The treatment of each of the intervals, $[W(k),W(k+1))$, begins by reconstructing the bit $1_{0<\text{ExclusiveInterval}(k)}$ from its secret shares.

If $\text{ExclusiveInterval}(k)=0$ (namely, if the interval is to be split among more than one agent, or if the interval is assigned to none) the subprotocol proceeds as follows. It maintains a secret-shared variable Position that stores the left end point of the next portion of the interval to be assigned to an agent; it is initialized to the left end point of the interval. Then, the subprotocol scans all agents, $i \in [n]$.
If $\text{IntervalAllocation}(i,k)>0$
the subprotocol computes the portion of that interval to be assigned to $A_i$ by computing the fraction in Eq.~(\ref{fraction}), using
the secure MPC division protocol.
After computing a secret sharing of AllocationSize---the length of the portion of the $k$-th interval allocated to $A_i$---the subprotocol proceeds to compute secret sharings of the two end points of that portion. If, however, $A_i$ was not assigned any portion of that interval, the two end points it would get are {\em obliviously} set to $Q$ (which stands for the right end point of the cake) so that the portion it gets is empty. 
At the end, 
all agents send their shares of those two values to $A_i$ who proceeds to reconstruct them and deduce the actual portion on the interval $[0,1)$ that it was given. Finally, the secret-shared value Position is moved forward by AllocationSize. 

Next, we discuss the case $\text{ExclusiveInterval}(k)>0$ in which the $k$-th interval is assigned as a whole to a single agent, say $A_i$. Assume that $A_i$ is assigned a sequence of consecutive whole intervals, say $[W(k),W(j))$ where $k<j \leq m$. A simple treatment of that case would send $j-k$ independent messages to $A_i$, one for each of the intervals $[W(h),W(h+1))$, $k \leq h < j$, that were assigned to it.
However, such an approach is inefficient, as it breaks a single portion, $[W(k),W(j))$, into $j-k$ intervals. Moreover, if $A_i$ is assigned the entire portion $[W(k),W(j))$, there is no need to expose the internal points $W(h)$, $k<h<j$, which correspond to private valuations of the agents. 
Hence, Subprotocol \textsc{FinalServing} is designed to generate a single message for $A_i$ for the entire portion $[W(k),W(j))$.

Note that the subprotocol sends a full set of $n$ messages to all agents, for each interval (or a block of consecutive intervals), even though most of those messages are redundant. That way,
\textsc{PP\_CC\_puv} maintains {\em restricted visibility}: namely, each agent is notified only of its own allocated portion, but remains oblivious regarding the allocations of its peers. In settings where a {\em full visibility} is desired, 
Subprotocol \textsc{FinalServing} can be modified so that instead of sending the shares of the two end points only to the designated agent, it will broadcast those shares, so that all agents get notified of the portions given to each one of them. 

\ignore{
Protocol \ref{wholep} outlines the entire cake cutting protocol.

\SetAlgorithmName{Protocol}{}{}
\begin{algorithm}[htb!]
\SetAlgoLined
\DontPrintSemicolon
{\bf \textsc{SharingPrivateValuations}} (Protocol \ref{Dist_Val})\\
{\bf \textsc{Intervals}} (Protocol \ref{Set_Intervals})\\
{\bf \textsc{IterativeAllocation}} (Protocol \ref{IterativeCode})\\
{\bf \textsc{FinalServing}} (Protocol \ref{Protocol_distribute_cake})\\
\caption{{\bf A privacy-preserving protocol for fair cutting of a cake}}
\label{wholep}
\end{algorithm}
}

\section{Properties}\label{properties}

\textsc{PP\_CC\_puv} offers perfect privacy and is strategy-proof. It preserves \textsc{CC\_puv}'s asymptotic runtime complexity and incurs only a quadratic increase in communication complexity. Those properties are stated in Theorems \ref{thm1}-\ref{thm3}. The proofs are given in Appendix B.

\begin{theorem}\label{thm1} (Perfect Security) If the set of agents has an honest majority then Protocol \textsc{PP\_CC\_puv} is perfectly secure.
\end{theorem}

\begin{theorem}\label{thm2} (Strategyproofness) Protocol \textsc{PP\_CC\_puv} is strategyproof for agents with piecewise uniform valuations.
\end{theorem}

\begin{theorem}\label{thm3} (Overheads)
The overheads incurred by \textsc{PP\_CC\_puv} on top of \textsc{CC\_puv} increase the overall runtime by at most a constant multiplicative factor, while the added communication complexity is at most $\mathcal{O}(n^2)$.
\end{theorem}

\section{Conclusion}
We presented a private, strategyproof, and envy-free cake-cutting protocol, building on the non-private algorithm of \citet{ChenLPP10,ChenLPP13}. Privacy is a critical consideration in many real-world allocation scenarios; without it, agents may withhold or distort their true preferences out of concern for exposure. By preserving strategyproofness within a privacy-preserving framework, our protocol enables agents to safely and truthfully report their valuations. As a result, the algorithm's envy-freeness holds with respect to the agents' actual preferences. In contrast, an envy-free algorithm that does not guarantee truthful reporting may produce allocations that appear envy-free based on misreported inputs, but fail to be so under the agents’ true valuations.

This work represents a first step toward establishing privacy guarantees for cake-cutting algorithms. We believe the methodology introduced here can serve as a foundation for privatizing a broader class of cake-cutting algorithms. While privacy is particularly powerful when combined with strategyproofness---addressing both incentive and confidentiality concerns---it also holds value in non-strategyproof settings. This is especially relevant in environments with non-strategic agents, such as automated systems that are restricted from misreporting, where the absence of privacy alone may discourage participation, regardless of strategic considerations.

\clearpage

\bibliography{refs}

\clearpage
\appendix

\section{Appendix}
The appendix contains additional discussions that could not be included in the main body of the paper due to space limitations. These materials are provided for the reviewers' consideration. We emphasize that the main text offers a comprehensive overview of our privacy-preserving cake-cutting protocol, which can be understood and appreciated independently of the supplementary details presented here. 

\subsection{Secure division}\label{sec_div}
Here we present a secure division protocol.
We start by recapping the binary long division algorithm.
Assume that the binary representation of the numerator $u$ is $(u_{s-1},\ldots,u_0)_2$, where $s = \lceil \log_2 p \rceil$ is the number of bits required to represent field elements in $\FF_p$. Then Algorithm \ref{longdiv} performs a binary long division between $u$ and $v$. Protocol
\ref{seclongdiv} is a secure implementation of that algorithm. Because the algorithm and protocol are straightforward, we omit further explanation.

\SetAlgorithmName{Algorithm}{}{}
\begin{algorithm}[h!!!]
\SetAlgoLined
\DontPrintSemicolon
 \Input{$u,v \in \ZZ$, $u \in [0,p)$, $v \in (0,p)$; $u_{s-1},\ldots,u_0$ such that $u = (u_{s-1},\ldots,u_0)_2$}
 {
$q \gets 0$\\
$r \gets 0$\\
\ForAll{$j=s-1,\ldots,0$}{
$r \gets 2r + u_j$\\
\If{$r<v$}{
$q_j \gets 0$\\
} \Else {
$q_j \gets 1$\\
$r \gets r - v$\\
}
$q \gets 2q+q_j$
}
}
 \Output{The quotient $q=\lfloor \frac{u}{v} \rfloor$ and remainder $r=u ~\mbox{mod}~ v$}
\caption{Binary long division}\label{longdiv}
\end{algorithm}

\SetAlgorithmName{Protocol}{}{}
\begin{algorithm}[h!!!]
\SetAlgoLined
\DontPrintSemicolon
 \Input{$[[u]]$, $[[v]]$, $v \in (0,p)$, and $[[u_j]]$, $0 \leq j \leq s-1$, such that $u = (u_{s-1},\ldots,u_0)_2$}
 {
$[[ q ]] \gets [[ 0 ]]$\\
$[[r]] \gets [[0]]$\\
\ForAll{$j=s-1,\ldots,0$}{
$[[r]] \gets 2[[r]] + [[u_j]]$\\
$[[q_j]] \gets [[1]] - [[ 1_{r<v}]]$\\
$[[r]] \gets [[r]] - [[q_j]]\cdot [[v]]$\\
$[[q]] \gets 2[[q]]+[[q_j]]$
}
}
 \Output{A sharing of $q=\lfloor \frac{u}{v} \rfloor$ and of $r=u ~\mbox{mod}~ v$, as well as sharings of $q$'s bits, $q_{s-1},\ldots,q_0$}
\caption{Secure binary long division}\label{seclongdiv}
\end{algorithm}


\subsection{Subprotocol \textsc{SharingPrivateValuations}}\label{App-Prot1}
Here we describe Subprotocol \ref{Dist_Val} (\textsc{SharingPrivateValuations}) that prepares the private valuation functions for secret sharing (see Section \ref{preparing}).

The subprotocol starts by having all agents agree on an upper bound $L$ on the number of intervals $\ell_i$, $i \in [n]$, in the private valuation functions (line 1). 
Recall that the description of each of the valuation functions is given by $2\ell$ integers, see Eq. (\ref{boundaries_int}), where $\ell = \max_{i \in [n]} \ell_i$. Before the value of $\ell$ is revealed, each agent describes its valuation function by $2L$ (rather than $2\ell$) integers, in the same manner as in Eq. (\ref{boundaries_int}), where the last $2(L-\ell_i)$ integers are all $Q$. Hence, after agreeing on $L$, the agents distribute $(t,n)$-shares (with $t$ as in Eq. (\ref{tdef})) of the $2L$ integers that describe their valuation functions, and also of $\ell_i$ (lines 2-8).

Next, the agents verify that the distributed shares correspond to legal inputs; agents who distributed shares of illegal inputs are marked as cheaters.
In line 10 the agents check that the secret sharing of $\ell_i$, for each $i \in [n]$, corresponds to $\ell_i \leq L$. If not, the relevant agent is marked as a cheater by the bit $\text{Cheater}_i$. That bit is secret-shared so that at this point there is still no public shaming.

Then, in lines 11-18, the agents verify that the sequence of $\{a_{i,j},b_{i,j}: j \in [L] \}$ that $A_i$, $i \in [n]$, had secret shared, complies with Eq. (\ref{boundaries_int}). If cheating is detected, then $A_i$ is secretly marked as a cheater, but still no public shaming is executed, nor any measures are taken.

Afterwards, in lines 19-24, the agents check the consistency of the sequence
$\{a_{i,j},b_{i,j}: j \in [L] \}$ with the secret-shared $\ell_i$, for each $i \in [n]$. Specifically, all values in the sequence after the first $2\ell_i$ ones must equal $Q$. Submitting secret shares of inputs that fail to comply with that requirement will also result in marking the relevant agent as a cheater.

Next, the agents reconstruct the bits $\text{Cheater}_i$ for all $i \in [n]$. If any cheaters are identified during the preceding three checks, their identities are revealed and the subprotocol aborts (lines 25–32). Otherwise, the subprotocol continues by computing the actual value of $\ell$ (lines 34–35), using the secure maximum MPC protocol (Section \ref{MPC}), and discards all secret shares associated with indices $j > \ell$, as they are no longer required.

The computation of $\ell$ in Subprotocol \ref{Dist_Val} serves only to improve efficiency by avoiding redundant computation and storage. Although $\ell$ reveals some information about the private valuations, that information is minimal and arguably benign. In application scenarios where even such minor information disclosure might be frowned upon, lines 33–36 can be safely omitted. The subprotocol can then proceed using the inflated representation of private valuations as initially set in lines 2–8. 

As for the handling of detected cheaters, it falls outside the scope of this work. Honest agents may choose to issue warnings and allow the cheaters to participate in a rerun of the subprotocol, or alternatively, they may opt to exclude them entirely. In the remainder of this discussion, we assume that Subprotocol~\ref{Dist_Val} has completed successfully. 

We conclude by describing a mechanism for jointly selecting the parameter \( Q \) at the outset of the subprotocol. $Q$ governs the precision of the mapping from the actual support boundaries of the valuations on \( C = [0,1] \), Eq.~\eqref{boundaries}, to their integer representations, Eq.~\eqref{boundaries_int}. A convenient choice is \( Q = 10^d \), where \( d \) corresponds to the number of significant decimal digits preserved by this mapping. Each agent \( A_i \), for \( i \in [n] \), knows the number of decimal digits \( d_i \) required to accurately encode its own valuation boundaries. The agents can then securely compute \( d = \max_{i \in [n]} d_i \), using a procedure analogous to their computation of \( \ell \) in line~34 of Subprotocol~\ref{Dist_Val}, and finally set \( Q = 10^d \).

\SetKwInput{Input}{Input}
\SetKwInput{Output}{Output}
\SetKwInput{Parameter}{Parameter}

\SetAlgorithmName{Subprotocol}{}{}
\begin{algorithm}[h!!!]
\SetAlgoLined
\DontPrintSemicolon
\Input{$A_i$, $i \in [n]$, has $\{a_{i,j},b_{i,j}:j \in [\ell_i]\}$, in compliance with Eq. (\ref{boundaries_int})}
 \Parameter{$~Q$ -- the discretization parameter; $~ t = \lfloor \frac{n+1}{2} \rfloor$ -- the secret sharing threshold}
 {
All agents agree on $L$, an upper bound on $\ell_i$, $i \in [n]$\\
\ForAll{$i \in [n]$}{\
\ForAll{$\ell_i < j \leq L$}{
$A_i$ sets $a_{i,j},b_{i,j} \gets Q$\\
}
\ForAll{$1 \leq j \leq L$}{
$A_i$ runs $\text{Share}_{t,n}(a_{i,j})$\\
$A_i$ runs $\text{Share}_{t,n}(b_{i,j})$\\
}
$A_i$ runs $\text{Share}_{t,n}(\ell_{i})$ \\
}
\ForAll{$i \in [n]$}{
$[[\mbox{Cheater}_i]] \gets [[1_{L<\ell_i}]]$\\
$[[b_{i,0}]] \gets [[0]]$\\
$[[a_{i,L+1}]] \gets [[Q]]$\\
\ForAll{$ 1 \leq j \leq L+1$}{
$[[\alpha_{i,j}]] \gets [[1_{b_{i,j-1} \leq a_{i,j} }]]$\\
}
\ForAll{$ 1 \leq j \leq L$}{
$[[\beta_{i,j}]] \gets [[1_{a_{i,j} \leq b_{i,j} }]]$\\
}
$[[\gamma_i]] \gets \prod_{j=1}^{L+1} [[ \alpha_{i,j} ]] \cdot \prod_{j=1}^{L} [[ \beta_{i,j}]]$\\
$[[\mbox{Cheater}_i]] \gets [[\mbox{Cheater}_i]] \vee [[1_{\gamma_i=0} ]]$\\
$[[\gamma_i]] \gets [[0]]$\\
\ForAll{$ 1 \leq j \leq L$}{
$[[\alpha_{i,j}]] \gets [[ 1_{a_{i,j}=Q}]] \cdot [[1_{b_{i,j}=Q}]] $\\
$[[\beta_{i,j}]] \gets [[ 1_{j\leq \ell_i}]]$\\
$[[ \gamma_i ]] \gets [[\gamma_i]] + ([[\alpha_{i,j}]] \vee [[\beta_{i,j}]])$\\
}
$[[\mbox{Cheater}_i]] \gets [[\mbox{Cheater}_i]] \vee [[1_{\gamma_i<L} ]]$\\
}
$\mbox{Abort} \gets 0$\\
\ForAll{$i \in [n]$}{
$\mbox{Cheater}_i \gets \text{Reconstruct}_t([[\mbox{Cheater}_i]])$\\
\If {$\textnormal{Cheater}_i = 1$} {
Mark $A_i$ as a cheater\\
$\mbox{Abort} \gets 1$
}
}

\If{$\textnormal{Abort} = 1$}{
 Abort Execution\\
}
\Else{
$[[\ell]] \gets \max([[\ell_i]]: i \in [n]) $\\
$\ell \gets \text{Reconstruct}_t([[\ell]])$\\
The agents drop received shares of $a_{i,j}$ and $b_{i,j}$ for all $i \in [n]$ and $\ell<j\leq L$\\
}
 \Output{A verified secret sharing of the valuations of all honest agents}
\caption{\textsc{SharingPrivateValuations}: Distributing shares of the private valuations}\label{Dist_Val}
}
\end{algorithm}

\subsection{Subprotocol \textsc{Intervals}}\label{intervalsprotocol}

Here we present Subprotocol \ref{Set_Intervals} (\textsc{Intervals}) that summarizes the interval computations as discussed in Sections \ref{sec_split} and \ref{intsec}.

\SetAlgorithmName{Subprotocol}{}{}
\begin{algorithm}[h!!!]
\SetAlgoLined
\DontPrintSemicolon
\Input{Secret sharings of $a_{i,j}$ and $b_{i,j}$ for all $i \in [n]$, and $j \in [\ell]$}
 {
\ForAll{$i \in [n]$}{
\ForAll{$j \in [\ell]$}{
$[[V(2(i-1)\ell+2j-1)]] \gets [[a_{i,j}]]$\\
$[[V(2(i-1)\ell+2j)]] \gets [[b_{i,j}]]$\\
}
}
$m \gets 2\ell n + 2$\\
$[[V(m-1)]] \gets [[0]]$\\
$[[V(m)]] \gets [[Q]]$\\
\ForAll{$k \in [m]$}{
$[[W(k)]] \gets$ The secret sharing of the $k$-th ranked element in $V$ (Using \citet{MPCoverSS} algorithm)\\
}
\ForAll{$k \in [m-1]$}{
$[[ \mbox{IntervalLen}(k)]] \gets [[ W(k+1) ]] - [[ W(k) ]]$\\
$[[ \mbox{IntervalAvailable}(k) ]] \gets [[1]]$\\
}
\ForAll{$i \in [n]$}{
$ [[\mbox{AllocationDenominator}(i)]] \gets [[0]]$\\
\ForAll{$k \in [m-1]$}{
Compute $[[\mbox{IntervalDesired}(i,k)]]$ by Eq. (\ref{postp})\\
$[[\mbox{IntervalAllocation}(i,k)]] \gets [[0]]$\\
}
}
 \Output{Secret sharings of the arrays $W$, IntervalLen, IntervalAvailable, IntervalDesired, AllocationDenominator, IntervalAllocation}
\caption{\textsc{Intervals}: Splitting the cake to intervals}\label{Set_Intervals}
}
\end{algorithm}

\subsection{Subprotocol \textsc{IterativeAllocation}}\label{sec:iterative}
Subprotocol~\ref{IterativeCode}, termed \textsc{IterativeAllocation}, finds a non-empty subset $S' \subseteq [n]$ that minimizes $\mbox{avg}^*(S',X)$, where $X$ is the subset of available intervals.
As discussed in Section \ref{basic_alg}, such a subset will consist only of yet unserved agents and it would also minimize $\mbox{avg}(S',X)$.

In the baseline version presented here, the subprotocol exhaustively scans all \(2^n - 1\) non-empty subsets of \([n]\), comparing the \(\mbox{avg}^*\) value of each subset to that of the best one found so far. We later describe a more sophisticated version that identifies the minimizing subset \(S'\) in time polynomial in \(n\).

In order to scan all such subsets, the subprotocol maintains a clear bit vector $\text{AgentsInSubset}(1:n)$, and it scans all $2^n-1$ non-zero assignments to that vector, where each assignment corresponds to a distinct subset $S' \subseteq [n]$. In addition, the subprotocol maintains secret sharings of the following values:
\begin{itemize}
 \item $\mbox{BestSubset}(1:n)$---the vector $\mbox{AgentsInSubset}(1:n)$ corresponding to the subset $S'$ that achieved a minimal value of $\mbox{avg}^*$ so far in the search.
 \item $\mbox{MinLen}$---the value of $\mbox{Len}^*$ of that subset.
 \item $\mbox{SizeBestSubset}$---the size of that subset.
\end{itemize}
At the outset of the search loop, the secret sharings of those variables are initialized as follows:
\be
\begin{aligned}
[[\text{BestSubset}(i)]] &\gets [[0]] \,, ~~~ i \in [n] \,, \\
[[\text{MinLen}]] &\gets [[n \cdot (Q+1)]] \,, \\
[[\text{SizeBestSubset}]] &\gets [[0]] \,.
\end{aligned}
\label{initvars}
\ee
Namely, the minimizing subset is initialized to the empty set with a value of $n\cdot(Q+1)$ to $\mbox{Len}^*$. Then, the subprotocol starts scanning all $2^n-1$ non-empty subsets of $[n]$, where the identity of the currently considered subset $S'$ is known to all through the public bit vector $\mbox{AgentsInSubset}(1:n)$.
In order to determine whether the current subset $S'$ in the search loop issues a smaller value to $\mbox{avg}^*$ than the best subset found so far, the agents need to check whether
$$ \frac{ \mbox{Len}^*(S',X)}{|S'|} < \frac{\mbox{MinLen}}{\mbox{SizeBestSubset}} \,,$$
or, alternatively, whether
$$ \mbox{SizeBestSubset} \cdot \mbox{Len}^*(S',X) <
|S'| \cdot \mbox{MinLen}
\,.$$
The agents maintain a secret sharing of $\mbox{SizeBestSubset}$ and of $\mbox{MinLen}$. As for $\mbox{Len}^*(S',X) $, they can compute secret shares of that value by Eqs. (\ref{lendef}),(\ref{hdef}) and (\ref{LenStar}). Finally, $|S'|$ is a public value as it equals $ \sum_{i \in [n]} \mbox{AgentsInSubset}(i)$. Hence, using secure multiplication and secure comparison (Section \ref{MPC}), the agents can compute a secret sharing of the bit 
\be \mbox{Update} \hskip -0.06cm := \hskip -0.06cm 1_{ \mbox{SizeBestSubset} \cdot \mbox{Len}^*(S',X) <
|S'| \cdot \mbox{MinLen}} \,. \label{updatebit}\ee
Then,
throughout the search for the subset $S'$ that minimizes
$\mbox{avg}^*$, the subprotocol obliviously updates the secret sharings of $\mbox{BestSubset}(i)$, $i \in [n]$, $\mbox{MinLen}$, and $\mbox{SizeBestSubset}$ as follows:
\be 
\begin{split}
[[\mbox{BestSubset}(i)]] \gets [[\mbox{Update}]] \cdot \mbox{AgentsInSubset}(i) \hskip 0.25cm \\
+ ([[1]] - [[\mbox{Update}]]) \cdot [[\mbox{BestSubset}(i)]] \,,
\end{split} 
\label{update1}\ee
\be
\begin{split}
[[\mbox{MinLen}]] \gets [[\mbox{Update}]] \cdot [[\mbox{Len}^*(S',X)]] \hskip 1cm \\ + ([[1]] - [[\mbox{Update}]]) \cdot [[\mbox{MinLen}]] \,,
\end{split}
\label{update2}
\ee
and
\be
\begin{split}
[[\mbox{SizeBestSubset}]] \gets [[\mbox{Update}]] \cdot |S'| \hskip 1.5cm \\ + ([[1]] - [[\mbox{Update}]]) \cdot [[\mbox{SizeBestSubset}]]
 \,.
\end{split}
\label{update3}\ee
As can be seen in Eqs. (\ref{update1})--(\ref{update3}), the secret shares of
$\text{BestSubset}(1:n)$, MinLen and SizeBestSubset are updated only if $\text{Update}=1$, and the update is done in an oblivious manner.

At the completion of the loop over all $2^n-1$ subsets, the agents hold secret shares of the vector $\mbox{BestSubset}(1:n)$ that describes the best subset $S'$ to serve in the current iteration of the subprotocol, without learning any information on that subset. The main computational task now is to perform an allocation of all cake intervals that are desired by at least one agent in $S'$ among the agents in $S'$. That is the topic of Section \ref{Step3}.

It remains now to update the two vectors $\mbox{AgentsServed}(1:n)$
and $\mbox{IntervalAvailable}(1:m-1)$. The secret shares of the first vector can be updated locally as follows:
\be \begin{split} [[\mbox{AgentsServed}(i)]] \gets [[\mbox{AgentsServed}(i)]] \hskip 1.4cm \\ +
[[\mbox{BestSubset}(i)]] 
\,, ~~~i \in [n]\,. \end{split} \label{partsrvd}\ee 
As for the second vector, the secret shares of each of its entries are updated as follows.
First, the agents compute secret shares of an array of secret bits, $\mbox{SelectedInterval}(1:m-1)$, in which the $k$-th bit indicates whether the $k$-th interval was selected in this iteration. The secret sharing of the $k$-th bit, $k \in [m-1]$, is computed by
\be \begin{split} 
[[\mbox{SelectedInterval}(k)]] \gets [[\mbox{IntervalAvailable}(k)]] \hskip 0.5cm\\
\cdot \bigvee_{i \in [n]} [[\mbox{BestSubset}(i)]]
\cdot [[\mbox{IntervalDesired}(i,k)]] 
\,. 
\end{split}\label{upd4t}\ee
Indeed, the bit $\mbox{SelectedInterval}(k)$ is set to 1 iff the $k$-th interval is available and desired by at least one agent in the found best subset $S'$ (that is described by the secret-shared vector $\mbox{BestSubset}(1:n)$). Then, the agents update the secret sharings of the bits $\mbox{IntervalAvailable}(k)$, $k \in [m-1]$, as follows: 
\be \begin{split}
[[\mbox{IntervalAvailable}(k)]] \gets [[\mbox{IntervalAvailable}(k)]] \hskip 0.3cm \\
\cdot \left( [[1]] - [[\mbox{SelectedInterval}(k)]] \right) \,.
\end{split}
\label{upd4}\ee
Namely, if an interval was already unavailable at the beginning of the iteration (since it had been served to one of the agents prior to that iteration), then
$\mbox{IntervalAvailable}(k)$ will remain 0. However, if it was available then its value will change from 1 to 0 if it was selected in the current iteration, as indicated by 
$\mbox{SelectedInterval}(k)$, Eq. (\ref{upd4t}).
The updates in Eqs. (\ref{partsrvd})--(\ref{upd4}) are executed before the computation in Section \ref{Step3} starts.

The above-described subprotocol is exponential in $n$. We concentrate on that version of the subprotocol for the sake of clarity. In their preliminary publication~\citet{ChenLPP10} did not discuss the implementation of the search for the subset with minimal average demand. 
However, in the full version of their study \cite{ChenLPP13} they described a polynomial-time search procedure. 
In Appendix \ref{sec:pol}, we describe that polynomial-time procedure and explain how it can be privatized using the same cryptographic techniques employed elsewhere in this work.

\medskip
Subprotocol \ref{IterativeCode} summarizes the computations of the iterative process.
It starts by initializing the secret shares of the bit vector AgentsServed and the counter NumAgentsServed that will secretly hold the identities of all agents that had been served thus far, and their number (lines 1-3).

The main loop (lines 4-35) proceeds as long as the number of agents served is less than $n$. The condition in line 4 is verified as follows: the agents hold secret shares of the counter NumAgentsServed, while $n$ is public. Using the secure MPC equality protocol (Section \ref{MPC}), they compute secret shares of the bit that indicates whether NumAgentsServed equals $n$, and then they reconstruct that bit. The while loop proceeds as long as that bit equals 0.

Each iteration begins by initializing the secret sharings of the BestSubset bit array, MinLen and SizeBestSubset (see Eq. (\ref{initvars})). Then the subprotocol starts scanning all $2^n-1$ non-empty subsets of $[n]$ (lines 8-27). For each number $u \in [2^n-1]$ it constructs the bit vector AgentsInSubset$(1:n)$ that holds a 0-1 encoding of a non-empty subset, and its corresponding size into NumAgentsInSubset (lines 9-12). Here, $u_j$ denotes the $j$-th bit in the binary representation of $u$, where $u_0$ is the least significant bit.

If $S'$ is the currently considered subset of $[n]$, the agents compute into $[[h]]$ a secret sharing of $h(S')$, as defined in Eq. (\ref{hdef}) (line 13). Note that here AgentsServed$(1:n)$ is a secret-shared vector while AgentsInSubset$(1:n)$ is a public vector. Next, the agents update $[[h]]$ to hold a secret sharing of the bit that equals 1 iff the current subset $S'$ consists entirely of yet-unserved agents (line 14). 

The subprotocol then proceeds to compute $\mbox{Len}(S',X)$, and subsequently $\mbox{Len}^*(S',X)$ of the current subset $S'$, using Eqs. (\ref{lendef}) and (\ref{LenStar}) (lines 15-20). 
\JOURNAL{
The command $\mbox{OR}(\{[[b_i]]: i \in [n]\})$ invokes the protocol for computing ORs of secret-shared bits (see Protocol \ref{ORprotocol} in Section \ref{MPCprotocols1}).
}

Next, the subprotocol computes secret shares of the bit Update, Eq. (\ref{updatebit}) (lines 21-23) and then it obliviously updates the secret sharings of BestSubset$(1:n)$, MinLen and SizeBestSubset, by Eqs. (\ref{update1})-(\ref{update3}) (lines 24-27).

After the loop over all $2^n-1$ subsets completes, 
the agents update the secret sharing of the bit array AgentsServed$(1:n)$ and NumAgentsServed (lines 28-30), as well as the secret sharing of the bit array IntervalAvailable$(1:m-1)$, by Eqs. (\ref{upd4t}+\ref{upd4}) (lines 31-34).
Finally, the agents execute the Subprotocol
\textsc{AssignCakeToSelectedAgents}, which assigns cake pieces to each of the agents in the selected subset $S'$ (line 35), see Section \ref{Step3} and Appendix \ref{Step3-app}.

\smallskip
\paragraph{\bf A concluding remark on privacy.}
Subprotocol \textsc{IterativeAllocation} is not perfectly secure, as it reveals the number of iterations executed until all agents are served—namely, the number of iterations of the {\bf while} loop in lines 4–34. This information is not part of the intended output and, ideally, should remain hidden.
In practical MPC protocols, some leakage of auxiliary information is often tolerated, provided that the information is considered benign. The number of iterations required in this context may be considered as such. Nevertheless, if stricter privacy is desired, the loop can instead be executed for exactly $n$ iterations—the maximum possible when serving $n$ agents—padding with no-op (void) iterations once all agents have been served. Thus, by incurring a higher runtime cost, one can eliminate even this minor form of information leakage.

\SetAlgorithmName{Subprotocol}{}{}
\begin{algorithm}[t!!!]
\SetAlgoLined
\DontPrintSemicolon
{
\ForAll{$i \in [n]$}{
$[[\text{AgentsServed}(i)]] \gets [[0]]$\\
}
$[[\text{NumAgentsServed}]] \gets [[0]] $\\
\While{$ (~1-\textnormal{Reconstruct}_t( [[ 1_{\textnormal{NumAgentsServed}=n} ]]) ~)$}{
$[[ \text{BestSubset}(i) ]] \gets [[0]]$, $\forall i \in [n]$\\
$[[\text{MinLen}]] \gets [[n \cdot (Q+1)]] $\\
$[[ \text{SizeBestSubset} ]] \gets [[0]]$\\
\ForAll{$1 \leq u \leq 2^n-1$}{
$\text{NumAgentsInSubset} \gets 0$\\
\ForAll{$i \in [n]$}{
$\text{AgentsInSubset}(i) \gets u_{i-1}$\\
$\text{NumAgentsInSubset} \gets \text{NumAgentsInSubset} + u_{i-1} $\\
}
$[[h]] \gets \sum_{i \in [n]} \text{AgentsInSubset}(i) \cdot [[ \text{AgentsServed}(i) ]]$\\
$[[h]] \gets [[ 1_{h = 0}]]$\\
$[[\text{Len}]] \gets [[0]]$\\
\ForAll{$k \in [m-1]$}{
$[[\text{Desired}]] \gets \text{OR}( \{\text{AgentsInSubset}(i) \cdot [[\text{IntervalDesired}(i,k)]] : i \in [n] \})$\\
$[[\text{Desired}]] \gets [[\text{IntervalAvailable}(k)]] \cdot [[\text{Desired}]] \cdot [[\text{IntervalLen} (k) ]]$\\
$[[\text{Len}]] \gets [[\text{Len}]] + [[\text{Desired}]]$\\
}
$[[\text{LenStar}]] \gets [[\text{Len}]] \cdot [[h]] + n(Q+1) \cdot ([[1]] - [[h]])$\\
$[[a]] \gets [[\text{SizeBestSubset}]] \cdot [[\text{LenStar}]]$\\
$[[b]] \gets \text{NumAgentsInSubset} \cdot [[\text{MinLen}]]$\\
$[[\text{Update}]] \gets [[1_{a<b}]]$\\
\ForAll{$i \in [n]$}{
$[[\text{BestSubset}(i)]] \gets ([[1]] - [[\text{Update}]]) \cdot [[\text{BestSubset}(i)]] +
[[\text{Update}]] \cdot \text{AgentsInSubset}(i)$\\
}
$[[\text{MinLen}]] \gets ([[1]] - [[\text{Update}]]) \cdot [[\text{MinLen}]] + [[\text{Update}]] \cdot
 [[\text{LenStar}]]$\\
$[[\text{SizeBestSubset}]] \gets ([[1]] - [[\text{Update}]]) \cdot [[\text{SizeBestSubset}]]
+ [[\text{Update}]] \cdot \text{NumAgentsInSubset}$\\
} 
\ForAll{$i \in [n]$}{
$[[\text{AgentsServed}(i)]] \gets [[\text{AgentsServed}(i)]] + [[\text{BestSubset}(i)]]$\\
}
$[[\text{NumAgentsServed}]] \gets [[\text{NumAgentsServed}]] + [[\text{SizeBestSubset}]] $\\
\ForAll{$k \in [m-1]$}{
$[[\text{Selected}]] \gets \text{OR}( \{[[\text{BestSubset}(i)]] \cdot [[\text{IntervalDesired}(i,k)]] : i \in [n] \})$\\
$[[\text{SelectedInterval}(k)]] \gets [[\text{IntervalAvailable}(k)]] \cdot [[\text{Selected}]]$\\
$[[\text{IntervalAvailable}(k)]] \gets [[\text{IntervalAvailable}(k)]] \cdot ([[1]]-
[[\text{SelectedInterval}(k)]])$\\
}
\textsc{AssignCakeToSelectedAgents}\\
} 
} 
\caption{\textsc{IterativeAllocation}: The iterative process of assigning cake to agents}\label{IterativeCode}
\end{algorithm}


\subsection{Subprotocol \textsc{AssignCakeToSelectedAgents}}\label{Step3-app}
The privacy-preserving Subprotocol \textsc{AssignCakeToSelectedAgents} (Subprotocol \ref{sffa})
computes a fair allocation of intervals in $X$ to agents in the selected subset $S'$, without having those sets available in the clear. Instead of operating on the graph $G(S',X)$, it operates on a super-graph in which layer 2 has a node for each interval in $\cW$ and layer 3 has a node for each agent in $\cA$.
As for the edge capacities, they vary across iterations and they are carefully defined to match those in the underlying algorithm,
\textsc{CC\_puv}, for edges between intervals in $X$ and agents in $S'$, while all other capacities are set to zero. The capacities are secret-shared to conceal the topology of the actual subgraph, thereby preserving the privacy of $S'$, $X$, and the desirability relationships between agents and intervals.

The structure of the graph is as follows:

\begin{itemize}
\item Layer 2 has $m-1$ nodes, one for each cake interval in $\cW$, see Eq. (\ref{intervals}).
\item Layer 3 has $n$ nodes, one for each agent in $\cA$. 
\item
The capacity of the edge from the source node to the node of the interval $[W(k),W(k+1))$, $k \in [m-1]$, is $A_k:=\mbox{SelectedInterval}(k) \cdot \mbox{IntervalLen}(k)$. Namely, it is either the length of the interval, if the interval is available and desired by at least one agent in the found best subset $S'$, or zero otherwise.
\item
The capacity of the edge from the node of the interval $[W(k),W(k+1))$, $k \in [m-1]$, to the node of agent $A_i$, $i \in [n]$, is $B_{k,i}:=\mbox{IntervalLen}(k) \cdot \mbox{IntervalDesired}(i,k) \cdot \mbox{BestSubset}(i)$. Namely, it either equals the length of the $k$-th interval, if $A_i$ desires it and $A_i$ is part of the selected subset in the current iteration, or zero otherwise. 
\item There is an edge from each node $A_i$, $i \in [n]$, in layer 3 to the target node with a capacity that equals 
$$C_i:=\mbox{avg}(S',X)= \frac{\mbox{Len}(S',X)}{|S'|} = \frac{\mbox{MinLen}}{\mbox{SizeBestSubset}}\,.$$
Recall that MinLen stands for $\mbox{Len}(S',X)$ (namely, the sum of lengths of intervals from $X$ that are desired by $S'$), while SizeBestSubset is $|S'|$. Both values are kept secret and the agents hold secret sharings of them, as computed in Subprotocol \ref{IterativeCode}, \textsc{IterativeAllocation}.
\end{itemize}

Observe that all edges between layers 3 and 4 have the same capacity. This set of edges also includes those originating from nodes corresponding to agents in $\mathcal{A} \setminus S'$. Ideally, the capacities of these edges should be set to zero. However, the construction of $B_{k,i}$ ensures that agents in $\mathcal{A} \setminus S'$ receive no incoming flow, rendering the outgoing flow from such nodes effectively irrelevant. As a result, we can retain the uniform edge capacities from all agent nodes $A_i$, for $i \in [n]$, as described above, since the maximum-flow algorithm will never route flow through edges emerging from agents not in $S'$. This design choice avoids the need for computationally expensive operations to distinguish between agents in $S'$ and those in $\mathcal{A} \setminus S'$.

Since we are interested only in finding a maximum flow in the graph, we will multiply all edge capacities by
$\mbox{SizeBestSubset}$ in order to avoid divisions.
The notations $A_k$, $B_{k,i}$, $C_{i}$, $k \in [m-1]$ and $i \in [n]$, will relate hereinafter to the edge capacities after that multiplication, as we summarize below:
\be 
\begin{aligned}
A_k &= \text{SizeBestSubset} \hskip -0.05cm \cdot \hskip -0.05cm\mbox{SelectedInterval}(k) \hskip -0.05cm \cdot \hskip -0.05cm\mbox{IntervalLen}(k) \\ 
B_{k,i} &= \text{SizeBestSubset} \hskip -0.05cm \cdot \hskip -0.05cm \mbox{IntervalLen}(k) \\ & \hskip 2.3cm \cdot \hskip -0.05cm \mbox{IntervalDesired}(i,k) \hskip -0.05cm \cdot \hskip -0.05cm\mbox{BestSubset}(i) \\
C_i &= \text{MinLen} 
\end{aligned}
\label{abcij}
\ee
Recall that the agents hold secret shares of each of the multiplicands on the right of Eq. (\ref{abcij}), so, by applying secure multiplication, they can also get secret sharings of the capacities $A_k$, $B_{k,i}$ and $C_i$, $k \in [m-1]$, $i \in [n]$.
With those capacities, the maximum flow in the graph will be 
\be \text{MaxFlow}=\text{MinLen} \cdot \text{SizeBestSubset}\,, \label{maxflow}\ee
since any maximal flow will route a flow of capacity MinLen out of each of the SizeBestSubset nodes corresponding to agents in $S'$.

\medskip
To find a maximum flow in the graph we apply the Ford-Fulkerson method 
\cite{Cormen}. In our case, since the graph is acyclic and all source-to-target paths have the same number of edges (three), the Ford-Fulkerson method becomes a simple greedy algorithm, as described in Algorithm \ref{ffa}. 
The algorithm maintains a corresponding set of variables,
$a_k$, $b_{k,i}$, $c_{i}$, where $k \in [m-1]$ and $i \in [n]$, that will hold the found maximum flow. All those variables are initialized to 0 (lines 1-6).
The variable flow will maintain the overall flow that the algorithm managed to push through the graph and it is also initialized to 0 (line 7).

Then, a greedy loop begins (lines 8–19), where the algorithm repeatedly searches for source-to-target paths with positive residual capacity. It scans all such paths (lines 9–10), and when it finds one with a positive minimal capacity (lines 11–12), it pushes that capacity along the path (line 13), increases the flow accordingly (lines 14–16), and updates the residual capacities (lines 17–19).

\SetAlgorithmName{Algorithm}{}{}
\begin{algorithm}[htb!]
\SetAlgoLined
\DontPrintSemicolon
 \SetKwInput{Input}{Input}
 \Input{$G(S',X)$ -- the graph with capacities $A_k,B_{k,i},C_i$, $k \in [m-1]$, $i \in [n]$;\\ $~~~~~~~~~~~~\mbox{MaxFlow}$ -- the expected maximum flow, as\\ $~~~~~~~~~~~~$given in Eq. (\ref{maxflow})}
 \ForAll{$k \in [m-1]$}{
 $a_k \gets 0$\\
 \ForAll{$i \in [n]$}{
 $b_{k,i} \gets 0$
 }
 }
\ForAll{$i \in [n]$}{
$c_i \gets 0$\\
}
$\mbox{flow} \gets 0$\;
\While {$\textnormal{flow} < \textnormal{MaxFlow}$} {
 \ForAll{$k \in [m-1]$}{
 \ForAll{$i \in [n]$}{
 $\mbox{MinCapacity} \gets \min(A_k,B_{k,i},C_i)$\;
 \If{$\textnormal{MinCapacity} > 0$}{
 $\mbox{flow} \gets \mbox{flow} + \mbox{MinCapacity}$\;
 $a_k \gets a_k + \mbox{MinCapacity}$\;
 $b_{k,i} \gets b_{k,i} + \mbox{MinCapacity}$\;
 $c_i \gets c_i + \mbox{MinCapacity}$\;
 $A_k \gets A_k - \mbox{MinCapacity}$\;
 $B_{k,i} \gets B_{k,i} - \mbox{MinCapacity}$\;
 $C_i \gets C_i - \mbox{MinCapacity}$\;
 }
 }
 }
}
 \SetKwInOut{Output}{Output}
 \Output{A maximum flow, described by $a_k$, $b_{k,i}$, $c_i$, $k \in [m-1]$, $i \in [n]$.}
\caption{Ford-Fulkerson algorithm for finding a maximum flow in the graph.}\label{ffa}
\end{algorithm}

Next, we devise a privacy-preserving implementation of Algorithm \ref{ffa}. As discussed earlier, the structure of the graph is publicly known, while the capacities of the edges are secret-shared.
The first part of Subprotocol \ref{sffa} (lines 1-19) emulates Algorithm \ref{ffa} by performing all computations on the secret sharings of the capacities.

\SetAlgorithmName{Subprotocol}{}{}
\begin{algorithm}[t!!!!]
\SetAlgoLined
\DontPrintSemicolon
 \SetKwInput{Input}{Input}
 \Input{$G(S',X)$ -- the graph with secret-shared capacities $[[A_k]],[[B_{k,i}]],[[C_i]]$, $k \in [m-1]$, $i \in [n]$, see Eq. (\ref{abcij});\\ $~~~~~~~~~~~\mbox{[[MaxFlow]]}$ -- a secret sharing of the expected maximum flow, see Eq. (\ref{maxflow})}
 \ForAll{$k \in [m-1]$}{
 $[[a_k]] \gets [[0]]$\\
 \ForAll{$i \in [n]$}{
 $[[b_{k,i}]] \gets [[0]]$
 }
 }
\ForAll{$i \in [n]$}{
$[[c_i]] \gets [[0]]$\\
}
$\mbox{[[flow]]} \gets [[0$]]\;
\While {$(~\textnormal{Reconstruct}_t([[1_{\textnormal{flow} < \textnormal{MaxFlow}}]]) ~)$} {
 \ForAll{$k \in [m-1]$}{
 \ForAll{$i \in [n]$}{
 $\mbox{[[MinCapacity]]} \gets [[\min(A_k,B_{k,i})]]$\;
 $[[\mbox{MinCapacity}]] \gets [[\min(\mbox{MinCapacity}, C_i )]]$\;
 $[[\mbox{flow}]] \gets [[\mbox{flow}]] + [[\mbox{MinCapacity}]]$\;
 $[[a_k]] \gets [[a_k]] + [[\mbox{MinCapacity}]]$\;
 $[[b_{k,i}]] \gets [[b_{k,i}]] + [[\mbox{MinCapacity}]]$\;
 $[[c_i]] \gets [[c_i]] + [[\mbox{MinCapacity}]]$\;
 $[[A_k]] \gets [[A_k]] - [[\mbox{MinCapacity}]]$\;
 $[[B_{k,i}]] \gets [[B_{k,i}]] - [[\mbox{MinCapacity}]]$\;
 $[[C_i]] \gets [[C_i]] - [[\mbox{MinCapacity}]]$\;
 }
 }
 }
\ForAll{$i \in [n]$}{
$[[\mbox{AllocationDenominator}(i)]] \gets [[\mbox{AllocationDenominator}(i)]] 
+ [[\mbox{SizeBestSubset}]] \cdot [[\mbox{BestSubset}(i)]]$\\ 
\ForAll{$k \in [m-1]$}{
$[[\mbox{IntervalAllocation}(i,k)]] \gets [[\mbox{IntervalAllocation}(i,k)]] + [[b_{k,i}]]$\\
}
}
 \SetKwInOut{Output}{Output}
\caption{\textsc{AssignCakeToSelectedAgents}: Allocating cake pieces to the agents in the selected subset.}\label{sffa}
\end{algorithm}

The maximum flow that Subprotocol \ref{sffa} (\textsc{AssignCakeToSelectedAgents}) computes in lines 1-19 determines the allocation of (a subset of the) intervals to (a subset of the) agents. That allocation is final in the sense that no agent in the current subset $S'$ will get any more cake portions later on. However, even though those allocations are part of the final output, they cannot be disclosed at this stage. Instead, they must be registered and disclosed together with all allocations from all iterations only after the while-iteration in Subprotocol \ref{IterativeCode} terminates. To that end, we use the secret-shared arrays
$ \mbox{IntervalAllocation}(1:n,1:m-1)$ and AllocationDenominator$(1:n)$ that we described in Section \ref{intsec} and initialized in Subprotocol \ref{Set_Intervals} there. 
The entry $ \mbox{IntervalAllocation}(i,k)$, $i \in [n]$, $k \in [m-1]$,
will register the portion of the $k$-th interval allocated to $A_i$, in any of the iterations.
After Subprotocol \ref{sffa} finds a maximum flow, it updates the relevant entries in the matrix, see line 23 in Subprotocol \ref{sffa}.
Since the value $b_{k,i}$ will be nonzero at most in one of the iterations, 
$\mbox{IntervalAllocation}(i,k)$ will equal, eventually, the part of the $k$-th interval that was allocated to $A_i$, multiplied by the size of the subset $S'$ to which $A_i$ belonged (that is $\mbox{SizeBestSubset}$, see Eq. (\ref{abcij})).
In order to get rid of the SizeBestSubset multiplier, we register its value in AllocationDenominator$(i)$, see line 21 in the subprotocol.
Since, for each $i \in [n]$, the bit $\mbox{BestSubset}(i)$ will be 1 only in one of the iterations, and in that iteration SizeBestSubset equals the size of the subset $S'$ in which $A_i$ was a member, the final value of
$\mbox{AllocationDenominator}(i)$ will be the correct denominator by which all
values $\mbox{IntervalAllocation}(i,k)$, $k \in [m-1]$, should be divided.


\subsection{Subprotocol \textsc{FinalServing}}\label{app:finalcut}
Here we describe Subprotocol~\ref{Protocol_distribute_cake}, referred to as \textsc{FinalServing}, that is executed after the iterative allocation had completed (Subprotocol~\ref{IterativeCode}, \textsc{IterativeAllocation}).
It computes the values $\text{Portion}(i,k)$, Eq.~(\ref{fraction}), which specify the fraction of the $k$-th interval allocated to agent $A_i$, for all $k \in [m-1]$ and $i \in [n]$. 

The subprotocol distinguishes between two types of intervals based on the number of agents that received a portion of them. 
Let \([W(k), W(k+1))\), for \(k \in [m-1]\), denote an interval, and let \(g(k)\) represent the number of agents that were allocated part of this interval, as determined by the secret-shared values \(\textnormal{IntervalAllocation}(i,k)\), which are updated by Subprotocol~\ref{sffa} (\textsc{AssignCakeToSelectedAgents}). 
Intervals for which \(g(k) = 1\) are considered \emph{exclusive}, meaning they are fully assigned to a single agent. Intervals for which \(g(k) \neq 1\) are \emph{non-exclusive}; they are either split among multiple agents or unassigned altogether. 
The loop in lines 1-8 of Subprotocol~\ref{Protocol_distribute_cake} determines the types of all $m-1$ intervals. For each interval, $[W(k),W(k+1))$, $k \in [m-1]$, the subprotocol checks whether agent $A_i$, $i \in [n]$, was assigned a portion of that interval (by updating the secret-shared bit IsAgentRelevant, line 5). In addition, the subprotocol updates the number of agents (NumberOfAgents) that were assigned a portion of that interval (line 6). In case $\text{NumberOfAgents}>0$, the secret-shared value RelevantAgent stores the index of the last such agent (line 7).
Finally, the secret-shared value $\text{ExclusiveInterval}(k)$ stores the index $i \in [n]$ of the agent that receives the whole interval, or 0 if that interval is to be split among $\text{NumberOfAgents}>1$ agents, or among none.

Next, the subprotocol scans all intervals and splits each one of them as indicated by the array $\text{IntervalAllocation}(1:n,1:m-1)$ and the array $\text{ExclusiveInterval}(1:m-1)$ (lines 9-34). First, the agents reconstruct the bit that indicates whether $\text{ExclusiveInterval}(k)>0$ or not. The setting of $\text{ExclusiveInterval}(k)$ in line 8 determines that if $\text{ExclusiveInterval}(k)>0$ then the $k$-th interval is assigned as a whole to the single agent $A_i$, where $i=\text{ExclusiveInterval}(k)$; however, if $\text{ExclusiveInterval}(k)=0$, the interval is split among more than one agent, or it is not allocated to any of the agents.

We begin with the case $\text{ExclusiveInterval}(k)=0$ (lines 11-22). The secret-shared variable Position will store the left end point of the next portion of the interval to be assigned to an agent, multiplied by $n!$ (that multiplication is executed in order to avoid fractions, as we explain below). The secret sharing of Position is initialized to the left end point of the interval, multiplied by $n!$ (line 12). Then, the subprotocol scans all agents (lines 13-21). First, it computes into IsAgentRelevant the bit that indicates whether that agent has been assigned a portion of the interval (line 14). It then computes into $\text{AllocationSize}$ a secret sharing of the fraction in Eq. (\ref{fraction}), multiplied by $n!$ (line 15). That computation is carried out by invoking the secure MPC division protocol (see Protocol \ref{seclongdiv} in Appendix \ref{sec_div}). That protocol takes secret sharings of two integers, $[[u]],[[v]]$, and returns secret sharings of their quotient, $q = \lfloor \frac{u}{v}\rfloor$, and remainder, $r = u \mod v$. By writing $\mbox{SecureDiv}_q$ in line 15 here we refer to the quotient output of Protocol \ref{seclongdiv}.
As can be seen in line 15, the numerator in the fraction is multiplied by $n!$; such a multiplication guarantees that dividing it by $\text{AllocationDenominator}(i)$ (which is an integer in $[n]$) issues a whole integer with no remainder.  
Hence, the computed $[[\text{AllocationSize}]]$ corresponds to the exact length of the portion to be given to $A_i$, multiplied by $Q$ (because of the $Q$-quantization) and by $n!$.

After computing a secret sharing of AllocationSize, the agents proceed to compute secret sharings of the two end points of that portion (lines 16-17). If $\text{IsAgentRelevant}=1$, the portion starts at Position and ends at Position+AllocationSize; however, if $\text{IsAgentRelevant}=0$ the subprotocol sets $\text{PortionStart},\text{PortionEnd} \gets n! Q$, a setting that indicates an empty portion. Subsequently, all agents send their shares of those two values to $A_i$ who proceeds to reconstruct them and deduce the actual portion on the interval $[0,1)$ (lines 18-20). Note that the actual portion on $[0,1)$ is obtained by dividing the two values
PortionStart and PortionEnd by $n! Q$, to compensate for the $Q$-quantization and the $n!$-rescaling.
Finally, Position is moved forward by AllocationSize (line 21). 

Lines 23-34 are dedicated to the case where the $k$-th interval is assigned as a whole to a single agent, say $A_i$. Assume that $A_i$ is assigned a sequence of consecutive whole intervals, say $[W(k),W(j))$ where $k<j \leq m$. A simple treatment of that case would send $j-k$ independent messages to $A_i$, one for each of the intervals $[W(h),W(h+1))$, $k \leq h < j$, that were assigned to it.
However, such an approach is inefficient, as it breaks a single portion $[W(k),W(j))$ into $j-k$ intervals. Moreover, if $A_i$ is assigned the entire portion $[W(k),W(j))$, there is no need to expose the internal points $W(h)$, $k<h<j$, which correspond to private valuations of the agents. While it is impossible to link any of those points to any of the private valuations that contributed it, our subprotocol is designed to prevent such unnecessary leakage of information, as it generates a single message for $A_i$ for the entire portion $[W(k),W(j))$.

The loop in lines 25-26 determines the boundaries of the sequence of intervals that are assigned, in their entirety, to the same agent. The agent who was assigned the $k$-th interval got the sequence of intervals from the $k$-th to the $(j-1)$-th. The following loop in lines 27-33 finds that agent, in an oblivious manner. It scans all agents $A_i$, $i \in [n]$, and computes a secret sharing of the bit IsAgentRelevant that indicates whether $A_i$ is the relevant agent. Then the subprotocol updates the secret shares of PortionStart and PortionEnd (lines 29-30) and generates messages to each of the agents. As in the previous case, the relevant agent will get the actual portion of the cake, while all other agents will get messages 
that indicate the empty portion $[1,1)$.

\medskip
We conclude this section with an example to consolidate our understanding of the process by which intervals are split among several agents. Consider a setting with $n=4$ agents and assume that $Q=100$ (i.e., all endpoints in all valuations require up to two digits after the decimal point).
Assume that the first ($k=1$) cake interval is $[0,0.2)$ and that it should be divided between $A_1$ and $A_2$, where $A_1$ is to receive a portion of length $0.125$ and $A_2$ a portion of length $0.075$. Finally, we assume that $A_1$ and $A_2$ are the only agents served in the same iteration, so that $\text{AllocationDenominator}(i)=2$ for $i=1,2$. 

Under the above assumptions, the incoming flow to the node representing the $k=1$ interval is $0.2 \cdot Q \cdot 2 = 40$, and the outgoing flows to the nodes of $A_1$ and $A_2$ are $25$ and $15$, respectively. By Eq. (\ref{fraction}), the fraction that represents the portions of the $k$-th interval allocated to $A_i$ is
$$ \mbox{Portion}(i,k):= \frac{\textnormal{IntervalAllocation}(i,k)}{\mbox{AllocationDenominator}(i)} \,.$$
In our case, $\mbox{Portion}(1,1)=25/2$ and $\mbox{Portion}(2,1)=15/2$.
In order to avoid dealing with fractions, we multiply those fractions by $n!=24$ and get
300 and 180, respectively. Consequently, in line 19, $A_1$ will recover $\text{PortionStart} = 0$ and $\text{PortionEnd} = 300$, while $A_2$ will recover $\text{PortionStart} = 300$ and $\text{PortionEnd} = 480$. These values will be translated in line 20, after dividing by $n! Q = 2400$, into the intervals $[0, 0.125]$ and $[0.125, 0.2]$, as required.

\subsection{Finding the Subset with Minimal Average Demand in Polynomial Time}\label{sec:pol}
As explained in Section~\ref{Sec:prelim}, in each iteration, \textsc{CC\_puv} considers the current subset~$S$ of unserved agents and the subset~$X$ of available cake intervals. It then identifies a subset~$S' \subseteq S$ that has the smallest average demand over~$X$, as defined in Eq.~(\ref{avgdef}). 
In Subprotocol \textsc{IterativeAllocation} (Section \ref{sec:iterative}) we assumed, for the sake of clarity, that the search for the subset $S' \subseteq S$ with minimal average demand is carried out by an exhaustive search, which is inefficient. Here we proceed to describe how that search can be done in a privacy-preserving manner in \emph{polynomial time}, based on the method outlined in \citet{ChenLPP13}.

Recall that in \textsc{PP\_CC\_puv}, all interval endpoints are quantized to integers in the range \([0\!:\!Q] := \{0, 1, \ldots, Q\}\), where \(Q = 10^d\) is chosen to ensure lossless quantization (see Section~\ref{preparing}). Also, recall that for any subset of agents \(S' \subseteq S\), the quantity \(\text{Len}(S', X)\) denotes the total length of intervals in \(X\) that are desired by agents in \(S'\) (see Section~\ref{basic_alg}). Hence, we have:
\[
\text{Len}(S', X) \in [0\!:\!Q]\,.
\]
It follows that the average demand, as defined in Eq.~(\ref{avgdef}), satisfies
\begin{align*}
\text{avg}(S', X) = \frac{\text{Len}(S', X)}{|S'|}
&\in [0\!:\!Q] / [n] \\
&:= \left\{ \frac{a}{b} \;\middle|\; a \in [0\!:\!Q],\, b \in [n] \right\}\,.
\end{align*}

Next, suppose we further rescale all interval endpoints by multiplying them by \(n!\), so that each endpoint becomes \(a \cdot n!\) for some \(a \in [0\!:\!Q]\). Under this transformation, we obtain:
\[
\text{avg}(S', X) \in F_{Q,n} := \left\{ \frac{a \cdot n!}{b} \;\middle|\; a \in [0\!:\!Q],\, b \in [n] \right\}\,.
\]
Thus, after \(Q\)-quantization and \(n!\)-rescaling, the value \(\text{avg}(S', X)\) is an integer belonging to the set \(F_{Q,n}\), which contains at most \(Qn + 1\) distinct values.

Consider the four-layer directed and capacitated graph \( G(S, X) \) constructed by \textsc{CC\_puv} for a given subset of agents \( S \) and a subset of intervals \( X \), as described in Section~\ref{Step3}. Let \( G(S, X; c) \) denote a variant of this graph, parameterized by an integer \( c \), which differs from \( G(S, X) \) only in the following way: in \( G(S, X; c) \), the capacities of all edges from the nodes in the third layer to the target node in the fourth layer are set to \( c \), instead of being set to \( \text{avg}(S, X) \) as in the original graph \( G(S, X) \).
\citet{ChenLPP13} proved the following lemma:
\begin{lemma}\label{lm2}
There is a flow of size $c|S|$ in \( G(S, X; c) \) iff for all $S' \subseteq S$, $\text{avg}(S',X) \geq c$.
\end{lemma}

For the sake of convenience, we shall refer hereinafter to a scalar $c$ as ``feasible" if the graph \( G(S, X; c) \) has a flow of size $c|S|$ and ``infeasible" otherwise. Then Lemma \ref{lm2} states that $c$ is feasible iff for all $S' \subseteq S$, $\text{avg}(S',X) \geq c$.

Let \( S_{\min} \subseteq S \) be a subset that minimizes \( \text{avg}(S', X) \) over all subsets \( S' \subseteq S \), and define \( c^* := \text{avg}(S_{\min}, X) \). By definition, we have \( \text{avg}(S', X) \geq c^* \) for all \( S' \subseteq S \).
It follows from the ``if'' direction of Lemma~\ref{lm2} that the graph \( G(S, X; c^*) \) has a flow of size \( c^* |S| \), namely, that $c^*$ is  feasible. Moreover, all \( c > c^* \) are infeasible. Indeed, if some $c>c^*$ was feasible, then by the ``only if'' direction of Lemma~\ref{lm2}, we would have \( \text{avg}(S_{\min}, X) \geq c \), contradicting the definition of \( c^* \).

\subsubsection{Computing $c^*$.}
Our first goal is to compute \( c^* = \text{avg}(S_{\min}, X) \). Since \( c^* \in F_{Q,n} \) and
all values in $F_{Q,n}$ are bounded by $Q \cdot n!$, we can find $c^*$ by a binary search, as described in Subprotocol \ref{c_star_binary_search}, named \textsc{CapacityBinarySearch}.
Given a candidate value $c$, it is needed to determine whether the graph $G(S,X;c)$ has a flow of size $c\cdot|S|$. In our case, since $S$ and $X$ are subsets that remain secret, 
Subprotocol \textsc{CapacityBinarySearch} is executed on the same fixed graph $G(\cA,\cW)$ as done also in
\textsc{AssignCakeToSelectedAgents} (see Section~\ref{Step3-app}). The differences between the graph that
\textsc{CapacityBinarySearch} examines and the graph that \textsc{AssignCakeToSelectedAgents} examines
are in the edge capacities, which, in both cases are all secret-shared.
There are two such differences: 

(a) While in \textsc{AssignCakeToSelectedAgents} all edge capacities are multiplied by {SizeBestSubset}, in \textsc{CapacityBinarySearch} they are all multiplied by \( n! \). In both cases, the multiplication is performed to ensure that all capacities are integers.

(b) In \textsc{CapacityBinarySearch}, the capacities of all edges between layer 3 and the target node are set to the provided tested value $c$, while in \textsc{AssignCakeToSelectedAgents} they are set differently (see Section~\ref{Step3-app}).

Subprotocol \textsc{CapacityBinarySearch} begins with setting the secret sharings of the initial lower and upper bounds of the search (lines 1,2). The lower bound is set to $c_L=0$, which is clearly feasible, while the upper bound of the search is set to $c_U=Q \cdot n!+1$, which is clearly infeasible. The search loop (lines 3-7) then starts. First, we set the secret sharing of the mid-value (line 4). 

The secret sharing on the right hand side of line 4 is of the value
$\lfloor x/2 \rfloor$ for $x=c_L + c_U$. 
Recall that since those values are secret-shared, they are embedded in the secret sharing finite field $\FF_p$ (see Section \ref{sss}). We proceed to explain how to compute
a secret sharing of $\lfloor x/2 \rfloor$, given a secret sharing of $x$.
First, we observe that $\lfloor x/2 \rfloor = x/2$ when $x$ is even, while
$\lfloor x/2 \rfloor = (x-1)/2$ when $x$ is odd.
Hence,
$ \lfloor x/2 \rfloor = \frac{p+1}{2} \cdot (x-\textnormal{LSB}(x) ) $, where $\frac{p+1}{2}$ is the inverse of $2$ in the finite field $\FF_p$, while $\textnormal{LSB}(x)$
is the least significant bit of $x$. The only ingredient missing for the computation in line 4 is 
the computation of a secret sharing of 
$\textnormal{LSB}(x)$ from a given secret sharing of $x$.
An MPC protocol that performs that computation was proposed in \cite{NO07}; interested readers may find there the details of that computation.

Then, the subprotocol determines whether the mid-value $c_M$ is feasible or infeasible by invoking the subprotocol \text{IsFeasible}, with the input $[[c_M]]$ (line 5).
It returns a secret sharing of 1 if $c_M$ is feasible and a secret sharing of 0 if $c_M$ is infeasible.

The subprotocol IsFeasible is a slight variation of Subprotocol \ref{sffa}. 
First, it looks for a maximum flow in $G(S,X;c_M)$, just like Subprotocol \ref{sffa} does. Then, it checks that the maximum flow equals $c_M \cdot |S|$. Since it computes a secret sharing of flow (the size of the maximum flow found) and also has secret sharings of $c_M$ (given as input) and of $|S|$, given by
$$[[~|S|~]] = [[n]] - \sum_{i\in [n]} [[\mbox{AgentsServed(i)}]]\,,$$
subprotocol IsFeasible may compute a secret sharing of $[[c_M]] \cdot [[~|S| ~]]$ by secure multiplication, and then compute a secret sharing in the desired output bit by securely testing equality between $[[c_M]] \cdot [[~|S|~]]$ and $[[\textnormal{flow}]]$.

Going back to Subprotocol \textsc{CapacityBinarySearch}, if $c_M$ is feasible ($b=1$) then $c_L \gets c_M$ while $c_U$ is kept the same; otherwise $c_L$ is kept the same while $c_U \gets c_M$ (lines 6-7). The loop proceeds that way until $c_U-c_L=1$ (line 3). Finally, the maximal feasible value of $c^*$ is $c_L$ (line 8). 

\medskip
Two comments on efficiency are in order: 
\begin{enumerate}
\item The number of iterations in \textsc{CapacityBinarySearch} is at most \(\lceil \log_2 (Q \cdot n! + 1) \rceil = \mathcal{O}(d + n \log n)\). However, since \(c^* \in F_{Q,n}\), and the set \(F_{Q,n}\) has size at most \(Qn + 1\), it is possible to perform the binary search over a sorted array of the elements in \(F_{Q,n}\), rather than over the full interval \([0, Q \cdot n! + 1]\). This optimization reduces the number of iterations to \(\mathcal{O}(d + \log n)\).
\item Let \( N \) denote the number of values over which \textsc{CapacityBinarySearch} performs binary search. The number of iterations in the while loop is always either \( \lfloor \log_2 N \rfloor \) or \( \lceil \log_2 N \rceil \). Therefore, instead of executing the termination condition check in every iteration (line~3), it is possible to perform \( \lfloor \log_2 N \rfloor \) iterations without checking the condition, and then verify the termination condition (namely, whether \( c_U - c_L = 1 \)) only once, in order to determine whether a final iteration is required.
\end{enumerate}

\SetAlgorithmName{Subprotocol}{}{}
\begin{algorithm}[t!!!!]
\SetAlgoLined
\DontPrintSemicolon
$[[c_L]] \gets [[0]]$\;
$[[c_U]] \gets [[Q \cdot n!+1]]$\;
\While{$(~\textnormal{Reconstruct}_t([[1_{c_U-c_L=1}]])=0 ~)$}{
$[[c_M]] \gets [[~ \lfloor (c_L + c_U)/2 \rfloor ~]]$\;
$[[b]] \gets \text{IsFeasible}([[c_M]])$\;
$[[c_L]] \gets [[b]] \cdot [[c_M]] + ([[1]] - [[b]]) \cdot [[c_L]] $\;
$[[c_U]] \gets [[b]] \cdot [[c_U]] + ([[1]] - [[b]]) \cdot [[c_M]]$\;
}
\Return $[[c_L]]$\;
\caption{\textsc{CapacityBinarySearch}: Computing a secret sharing of the average demand of the subset $S_{\min} \subseteq S$ over $X$}\label{c_star_binary_search}
\end{algorithm}

\subsubsection{Using $c^*$ to find $S_{\min}$.}
After obtaining (a secret sharing of) \( c^* \), which represents the average demand over the set \( X \) of the desired subset \( S_{\min} \), we proceed to describe how \( c^* \) can be used to identify \( S_{\min} \).
\citet{ChenLPP13} showed that the desired $S_{\min}$ can be identified by computing a minimum cut in \( G(S, X; c^* + 1) \). That computation can be carried out as follows.

First, we run Subprotocol~\ref{sffa} over \( G(S, X; c^* + 1) \) and compute a maximum flow in it. Recall that during its exeution, Subprotocol~\ref{sffa} keeps updating the capacities of edges through which flow is pushed. Hence, at its completion, we obtain the residual graph, defined as the graph \( G(S, X; c^* + 1) \) with edge capacities reduced by the amount of flow sent through each edge in the computed maximum flow. 
The subset \( S_{\min} \) is then identified as the set of agents for which the corresponding nodes are no longer reachable from the source node in that residual graph.
Specifically, the node in layer 3 corresponding to the $i$-th agent is not reachable from the source node iff in the residual graph we have
\be \sum_{k \in [m-1]} \min(A_k,B_{k,i}) =0\,, \label{noreach}\ee
where $A_k$ are the residual capacities of the edges from the source node to the $k$-th interval node in layer 2, and $B_{k,i}$ are the residual capacities of the edges from layer 2 to layer 3, $k \in [m-1]$, $i \in [n]$.

However, since our privacy-preserving protocol operates over a fixed complete graph \( G(\mathcal{A}, \mathcal{W}) \), which includes nodes for all agents---including those who were already served in previous iterations---the above condition is insufficient, as it may inadvertently identify agents who have already been served. Therefore, the secure protocol must eliminate such free-riders by additionally verifying that
\begin{equation}
\text{AgentsServed}(i) = 0\,. \label{stillthere}
\end{equation}
In conclusion, the desired subset \( S_{\min} \) consists of all agents \( i \in [n] \) for whom both Eqs.~(\ref{noreach}) and~(\ref{stillthere}) hold.

\onecolumn

\SetAlgorithmName{Subprotocol}{}{}
\begin{algorithm}[htb!]
\SetAlgoLined
\DontPrintSemicolon
{
\ForAll{$k \in [m-1]$}{
 $[[\text{NumberOfAgents}]] \gets [[0]]$\;
 $[[\text{RelevantAgent}]] \gets [[0]]$\;
 \ForAll{$i \in [n]$}{
 $[[\text{IsAgentRelevant}]] \gets [[1_{0<\textnormal{IntervalAllocation}(i,k)}]]$\;
 $[[\text{NumberOfAgents}]] \gets [[\text{NumberOfAgents}]] + [[\text{IsAgentRelevant}]]$\;
 $[[\text{RelevantAgent}]] \gets [[\text{IsAgentRelevant}]] \cdot i + ([[1]]-[[\text{IsAgentRelevant}]]) \cdot [[\text{RelevantAgent}]] $\;
 }
 $[[\text{ExclusiveInterval}(k)]] \gets [[1_{\text{NumberOfAgents}=1}]] \cdot [[\text{RelevantAgent}]]$\;
}
$k \gets 1$\\
\While{$k \leq m-1$}{
\If{$(~\textnormal{Reconstruct}_t([[1_{0<\textnormal{ExclusiveInterval}(k)}]])=0~)$}{
 $[[\text{Position}]] \gets n! \cdot [[W(k)]] $\;
 \ForAll{$i \in [n]$}{
 $[[\text{IsAgentRelevant}]] \gets [[1_{0<\textnormal{IntervalAllocation}(i,k)}]]$\;
 $[[\text{AllocationSize}]] \gets \text{SecureDiv}_q(n! \cdot [[\text{IntervalAllocation}(i,k)]], [[\text{AllocationDenominator}(i)]])$\;
 $[[\text{PortionStart}]] \gets [[\text{IsAgentRelevant}]] \cdot [[\text{Position}]]+$ $([[1]] - [[\text{IsAgentRelevant}]]) \cdot n! \cdot Q$\;
 $[[\text{PortionEnd}]] \gets [[\text{IsAgentRelevant}]] \cdot ([[\text{Position}]] + [[\text{AllocationSize}]]) + $ $([[1]] - [[\text{IsAgentRelevant}]]) \cdot n! \cdot Q$\;
 All agents send their shares of PortionStart and PortionEnd to $A_i$\\
 $A_i$ recovers PortionStart and PortionEnd\\
 $A_i$ learns that it got the piece of cake $[\frac{\text{PortionStart}}{n! Q},\frac{\text{PortionEnd}}{n! Q})$\\
 $[[\text{Position}]] \gets [[\text{Position}]] + [[\mbox{AllocationSize}]]$\\
 }
 $k \gets k+1$\\
 }
 \Else
 {
 $j \gets k+1$\;
 \While{$(~(j \leq m-1) \wedge( \textnormal{Reconstruct}_t([[1_{\textnormal{ExclusiveInterval}(j)=\textnormal{ExclusiveInterval}(j-1)}]])=1)~)$}{
 $j \gets j+1$\;
 }
 \ForAll{$i \in [n]$}{
 $[[\text{IsAgentRelevant}]] \gets [[1_{\text{ExclusiveInterval}(k)=i}]]$\;
 $[[\text{PortionStart}]] \gets [[\text{IsAgentRelevant}]] \cdot [[W(k)]] + ([[1]] - [[\text{IsAgentRelevant}]]) \cdot Q$\;
 $[[\text{PortionEnd}]] \gets [[\text{IsAgentRelevant}]] \cdot [[W(j)]] + ([[1]] - [[\text{IsAgentRelevant}]]) \cdot Q$\;
 All agents send their shares of PortionStart and PortionEnd to $A_i$\\
 $A_i$ recovers PortionStart and PortionEnd\\
 $A_i$ learns that it got the piece of cake $[\frac{\text{PortionStart}}{Q},\frac{\text{PortionEnd}}{Q})$\\
 }
 $k \gets j$\;
 }
}
\SetKwInOut{Output}{Output}
\Output{The portions revealed securely to the relevant agents}
\caption{\textsc{FinalServing}: Splitting the cake among the agents.}\label{Protocol_distribute_cake}
}
\end{algorithm}

\newpage
\twocolumn

\section{Proofs}\label{app:proofs}
We provide here the full proofs of Theorems \ref{thm1}-\ref{thm3} stated in Section~\ref{properties}.

\subsection{Proof of Theorem~\ref{thm1} (Perfect Security)}
We prove perfect security in the standard information-theoretic, honest-majority
ideal/real simulation paradigm, using the security model and guarantees
described in Section~\ref{model}.

\paragraph{Ideal functionality.}
Let $\mathcal{F}_{\textsc{PPCC}}$ denote the ideal functionality that:
(i) receives from each agent $A_i$ its piecewise-uniform valuation $v_i$, encoded in the specified format;  
(ii) computes the allocation output of the \textsc{CC\_puv} algorithm; and  
(iii) reveals to agents only the output pieces permitted by the visibility mode
(restricted visibility: each agent learns only its own piece; full visibility: all agents learn the full allocation).
No intermediate quantities or auxiliary information are leaked.

\paragraph{Real protocol.}
The real execution proceeds as follows:
1. Each agent secret-shares its valuation into Shamir $t$-out-of-$n$ shares, where $t=\lfloor (n+1)/2 \rfloor$, see Eq. (\ref{tdef}).  
2. The servers jointly evaluate all steps of the \textsc{CC\_puv} allocation algorithm over secret shares:
   comparisons, equality predicates, min/max operations, OR operations, threshold checks,
   multiplications, divisions, and linear arithmetic.
3. The computed allocation is revealed according to the selected visibility mode.

Only the outputs permitted by the ideal functionality are revealed in plaintext; all other intermediate values remain hidden at all times because they appear solely in secret-shared form.

\paragraph{Simulator.}
Fix any adversary $\mathcal{A}$ corrupting fewer than $t$ parties (servers and/or agents) in the semi-honest model assumed in Section~\ref{model}.
We construct a perfect simulator $\mathsf{Sim}$ that produces a simulated view
distributed identically to $\mathcal{A}$'s real view, given only the corrupted inputs and the outputs allowed by $\mathcal{F}_{\textsc{PPCC}}$.

\emph{Secret sharing.}
For every secret-shared value $s$, $\mathsf{Sim}$ samples uniformly random shares for the corrupted parties.
Since Shamir sharing with $<t$ shares is perfectly hiding, these shares are identically distributed as in the real world.

\emph{MPC subprotocols.}
All MPC primitives used—secure multiplication, comparison, equality, min/max, OR, and division—admit perfect simulators in the honest-majority setting.
In each call, $\mathsf{Sim}$ replaces all transcripts with those produced by the corresponding ideal simulator, using only ideal outputs and fresh randomness.
The resulting transcript is identically distributed to the real protocol.

\emph{Reveals.}
If restricted visibility is used, $\mathsf{Sim}$ reveals to each corrupted agent only the output piece $\mathcal{F}_{\textsc{PPCC}}$ authorizes.
If full visibility is used, $\mathsf{Sim}$ reveals the complete allocation.
No other information is revealed.

\paragraph{Composition.}
All building blocks are perfectly simulatable under honest-majority Shamir secret sharing.
By the perfect composition theorem for MPC~\cite{BGW88}, the joint protocol is also perfectly simulatable.
Hence the adversary’s real view is perfectly determined by its ideal output view, completing the proof.
$\hfill \Box$

\subsection{Proof of Theorem~\ref{thm2} (Strategyproofness)}
Algorithm \textsc{CC\_puv} was shown to be strategyproof for piecewise uniform valuations by \citet{ChenLPP10}, by analyzing what the agents can do in order to affect the selection of the subset $S'$ with minimal average demand during the iterative phase of the algorithm.
Protocol \textsc{PP\_CC\_puv} securely computes \emph{exactly} the same allocation as
\textsc{CC\_puv} on any reported profile.
The cryptographic layer hides all intermediate information but does not modify or distort any decision logic of the mechanism.
Thus, for any reported valuation profile $(\hat v_i,\hat v_{-i})$, the output of \textsc{PP\_CC\_puv} equals the output of \textsc{CC\_puv} on the same input.

Therefore, no agent can gain by reporting $\hat v_i \ne v_i$, because such deviations cannot improve outcomes in \textsc{CC\_puv}, and \textsc{PP\_CC\_puv} preserves exactly those outcomes.

However, \textsc{PP\_CC\_puv} includes a preliminary step, in which the agents distribute secret shares of the number of the disjoint intervals in their private valuations ($\ell_{i}$), after which the protocol proceeds to compute in a secure manner
$\ell = \max_{i \in [n]} \ell_i$, and discloses it.
We wish to show that this additional computation, that was not present in \textsc{CC\_puv}, does not open new opportunities for manipulations.

If an agent overreports by submitting a value $\ell_i'$ that is greater than the true value $\ell_i$, the computed maximum $\ell$ may increase, but this only enlarges the encoding length and cannot improve the allocation it receives.

If, on the other hand, an agent underreports ($\ell_i' < \ell_i$), 
it has no effect on other agents but it may harm $A_i$ itself. 
The value that the protocol will compute in wake of this misreporting is $\ell':=\max\{ \ell'_i, \ell_j : j \neq i\}$. Since $\ell' \geq \ell_j$ for all $j \neq i$, then any other agent would still be able to report its true valuation. But it is possible that $\ell'$ would be smaller than $\ell_i$. In such cases, $A_i$ will not be able to fully report its valuation, while everyone else can. Since the algorithm is strategyproof, this means that agent $A_i$ will not benefit from this manipulation.

Hence, the step in which $\ell$ is computed does not create beneficial deviations. Therefore, since \textsc{PP\_CC\_puv} implements the same allocation function as the truthful, strategyproof algorithm \textsc{CC\_puv}, and since no new profitable deviation is introduced by the encoding-length step, truthful reporting remains a dominant strategy.
$\hfill \Box$

\subsection{Proof of Theorem~\ref{thm3} (Overheads)}

The secure and perfect emulation of the \textsc{CC\_puv} algorithm comes with a price. Here we analyze the additional computational and communication costs introduced by the protocol \textsc{PP\_CC\_puv}, as compared to its plaintext counterpart \textsc{CC\_puv}.

\paragraph{Constant-factor runtime overhead.}
In the honest-majority MPC setting used by the protocol:
\begin{itemize}
\item additions and affine transformations are local operations on shares;
\item multiplications require a constant number of rounds~\cite{ChidaGHIKLN18};
\item comparisons, equality tests, min/max, OR, and division each have constant-round protocols~\cite{NO07}.
\end{itemize}
Therefore, each arithmetic or logical step of \textsc{CC\_puv} is replaced by a constant-round MPC subroutine, yielding only a constant multiplicative increase in runtime.

\paragraph{Asymptotics preserved via fixed oblivious graph.}
\textsc{CC\_puv} builds a flow network over agents and nonempty valuation intervals as they arise.
To prevent topology leakage, \textsc{PP\_CC\_puv} instead evaluates a fixed four-layer graph over all agents and all interval boundaries.
This change preserves the asymptotic size of the network used in each max-flow call.
Thus the asymptotic work of each step of \textsc{CC\_puv} is unchanged, modulo the constant MPC factors above.

\paragraph{Communication dominated by pairwise operations.}
Each MPC multiplication or comparison requires $O(n^2)$ messages (pairwise exchanges of shares).
The protocol executes asymptotically the same number of logical operations as \textsc{CC\_puv},
up to constant factors, so the total communication is $O(n^2)$.

\paragraph{Conclusion.}
Runtime increases by at most a constant factor because every operation of \textsc{CC\_puv} is emulated by a constant-round MPC primitive.
Communication increases to $O(n^2)$ overall due to pairwise message patterns in MPC.
$\hfill \Box$


\end{document}